\documentclass[12pt,draftcls,onecolumn]{IEEEtran}

\ifCLASSINFOpdf
\else
\fi

\ifCLASSOPTIONcompsoc
\usepackage[nocompress]{cite}
\else
\usepackage{cite}
\fi

\usepackage{amsmath}
\usepackage{amssymb}
\usepackage{graphicx}
\usepackage{epstopdf}
\usepackage{mathtools}
\usepackage{xcolor}

\begin{document}

\title{Omnidirectional Precoding and Combining Based Synchronization for Millimeter Wave Massive MIMO Systems}

\author{
Xin Meng, Xiqi Gao, and Xiang-Gen Xia


\thanks{X. Meng and X. Q. Gao are with the National Mobile Communications Research Laboratory, Southeast University, Nanjing, 210096, China (e-mail: \{xmeng, xqgao\}@seu.edu.cn). X.-G. Xia is with the Department of Electrical and Computer Engineering, University of Delaware, Newark, DE 19716, USA (e-mail: xxia@ee.udel.edu).}
}



\maketitle

\vspace*{-28pt}

\begin{abstract}
In this paper, we design the precoding matrices at the base station side and the combining matrices at the user terminal side for initial downlink synchronization in millimeter wave massive multiple-input multiple-output systems. First, we demonstrate two basic requirements for the precoding and combining matrices, including that all the entries therein should have constant amplitude under the implementation architecture constraint, and the average transmission power over the total $K$ time slots taking for synchronization should be constant for any spatial direction.  Then, we derive the optimal synchronization detector based on generalized likelihood ratio test. By utilizing this detector, we analyze the effect of the precoding and combining matrices to the missed detection probability and the false alarm probability, respectively, and present the corresponding conditions that should be satisfied. It is shown that, both of the precoding and combining matrices should guarantee the perfect omnidirectional coverage at each time slot, i.e., the average transmission power at each time slot is constant for any spatial direction, which is more strict than the second basic requirement mentioned above. We also show that such omnidirectional precoding matrices and omnidirectional combining matrices exist only when both of the number of transmit streams and the number of receive streams are equal to or greater than two. In this case, we propose to utilize Golay complementary pairs and Golay-Hadamard matrices to design the precoding and combining matrices. Simulation results verify the effectiveness of the propose approach.
\end{abstract}

\begin{IEEEkeywords}
\textit{Millimeter wave (mmWave), massive multiple-input multiple-output (MIMO), synchronization, Golay complementary pair, Golay-Hadamard matrix}
\end{IEEEkeywords}

\newpage


\section{Introduction}

In recent years, utilizing millimeter wave (mmWave) frequency bands from 30 GHz to 300 GHz for cellular wireless communications has received considerable interest from both academia and industry \cite{Z.Pi2011,T.S.Rappaport2013,M.R.Akdeniz2014,W.Roh2014,T.Bai2015}. Compared with the existing cellular systems operated at carrier frequencies below 6 GHz, mmWave frequencies can offer orders of magnitude more spectrum to support higher data rates. Moreover, the small mmWave wavelengths also make it practical to deploy massive antenna arrays at both of the base station (BS) and user terminal (UT) sides. Hence considerable directional beamforming gains can be provided to compensate for the high isotropic path loss under mmWave frequencies. Therefore, mmWave massive multiple-input multiple-output (MIMO) is considered as one of the major technologies for next generation cellular systems.

Initial synchronization, also referred to as cell search or cell discovery in some literature, is a basic prerequisite to cellular communications. Generally, a BS broadcasts downlink synchronization signals periodically and a UT utilizes these signals to keep time and frequency synchronization with the BS, and then payload data transmission can be established. In the current cellular systems such as long-term evolution (LTE), to ensure cell-wide coverage, a BS usually transmits downlink synchronization signals by using a fixed wide beam pattern \cite{X.Yang2013}. Only after the synchronization has been established and the correct beamforming directions have been obtained, directional narrow beams are used to provide beamforming gains to improve the data rates.

When considering mmWave massive MIMO systems, it was once mentioned that directional narrow beams should also be used in the initial synchronization stage, as well as in the data transmission stage, to overcome the high isotropic path loss. Otherwise, there will be the problem that at one certain distance between a BS and a UT, although a reasonable data rate can be achieved by using directional transmission with beamforming gains, the synchronization cannot be established by using a wide beam pattern with a very low beamforming gain \cite{T.S.Rappaport2015,Q.Li2015}. On the other hand, it is known that different from data signals, the transmitted downlink synchronization signals usually consist of a predefined sequence that is foreknown at both of the BS and UT sides. This may provide additional spreading gains to increase the range of coverage for synchronization signals. For example in LTE, it is a Zadoff-Chu (ZC) sequence of length $63$ \cite{S.Sesia2009}, and the corresponding spreading gain is about $18$ dB. Moreover, although directional transmission increases the range of coverage, it also increases the latency time of synchronization. This is because multiple narrow beams towards different directions have to be used at multiple time slots to guarantee omnidirectional coverage in an average sense, since the correct beamforming directions are not known in the initial synchronization stage.

There have been some studies on mmWave massive MIMO synchronization. The results in \cite{C.N.Barati2015,C.N.Barati2016} showed that omnidirectional transmission is better than random beamforming, and full digital architectures with low resolution has significant benefits in comparison with single-stream analog beamforming. In \cite{C.Liu2016}, the authors identified the desired beam pattern in a targeted detectable region and approximated this beam pattern with the proposed designs. The optimal beamforming vectors maximizing the signal-to-noise ratio (SNR) values under different implementation architecture constraints were investigated in \cite{V.Raghavan2016}. In \cite{L.You2017}, a per-beam synchronization approach was proposed to moderate the variance of the beam domain channel. In this paper, we mainly consider that when the latency time taking for synchronization is fixed, how to design the precoding matrix at the BS side and the combining matrix at the UT side, to optimize the synchronization performance. Our main contributions are as follows.

\begin{itemize}
\item We demonstrate two basic requirements for the precoding matrices at the BS side and the combining matrices at the UT side, including that all the entries therein should have constant amplitude to satisfy the implementation architecture constraint, and the average transmission power over the total $K$ time slots should be constant for any spatial direction, where $K$ corresponds to the total latency time taking for synchronization.
\item We derive the optimal synchronization detector based on generalized likelihood ratio test (GLRT). By utilizing this detector, we analyze the effect of the precoding and combining matrices to the missed detection (MD) probability under two special channel models, respectively, including the single-path channel and the independent and identically distributed (i.i.d.) channel, and analyze the effect to the false alarm (FA) probability. We also present the corresponding conditions that the precoding and combining matrices should satisfy. It is shown that, both of the precoding and combining matrices should guarantee the perfect omnidirectional coverage at each time slot, i.e., the average transmission power at each time slot is constant for any spatial direction, which is more strict than the second basic requirement mentioned above.
\item We show that to guarantee constant amplitude for all the entries in the precoding and combining matrices, and at the same time guarantee the perfect omnidirectional coverage, both of the number of transmit streams and the number of receive streams should be equal to or greater than two. In this case, we propose to use Golay complementary pairs and Golay-Hadamard matrices to design the precoding and combining matrices.
\end{itemize}

{\color{black}{Note that omnidirectional coverage for payload data transmission in traditional MIMO systems has been considered in \cite{X.Yang2013}, where the authors propose a single-stream solution for the precoding vector, and analyze the system performance in terms of ergodic capacity and bit error rate. In this paper, we mainly focus on initial downlink synchronization in mmWave MIMO systems. We use MD probability and FA probability to evaluate the
synchronization performance, and propose a multiple-stream solution for the precoding and combining matrices to guarantee perfect omnidirectional coverage.}}

The rest of this paper is organized as follows. The system model is presented in Section II, including the synchronization signal model and the channel model. The basic requirements of the precoding and combining matrices are
demonstrated in Section III. The effect of the precoding and combining matrices to synchronization performance is analyzed in Section IV. The precoding and combining matrices are designed in Section IV. Numerical results
are presented in Section V. Finally, conclusions are drawn in Section VI.

\emph{Notations:} We use upper-case and lower-case boldfaces to denote matrices and column vectors. ${\bf{I}}_M$, ${\bf{1}}_{M}$, and $\bf{0}$ denote the $M \times M$ identity matrix, the $M \times 1$ column vector of all ones, and the zero matrix with proper dimensions, respectively. $(\cdot)^*$, $(\cdot)^T$, and $(\cdot)^H$ denote the conjugate, the transpose, and the conjugate transpose, respectively. ${\mathbb{E}}(\cdot)$ refers to the expectation and ${\mathbb{P}}(\cdot)$ represents the probability. The Kronecker product of two matrices $\bf{A}$ and $\bf{B}$ is denoted by ${\bf{A}} \otimes {\bf{B}}$. ${{[ {\bf{A}} ]_{m,n}}}$, ${{[ {\bf{A}} ]_{m,:}}}$, and ${{[ {\bf{A}} ]_{:,n}}}$ denote the $( {m,n} )$th element, the $m$th row vector, and the $n$th column vector of matrix $\bf{A}$, respectively, and ${{[ {\bf{a}} ]_m}}$ denotes the $m$th element of vector $\bf{a}$. ${\rm{diag}}( {\bf{a}} )$ and ${\rm{diag}}( {\bf{A}} )$ denote the diagonal matrix with $\bf{a}$ on the main diagonal and the column vector constituted by the main diagonal of $\bf{A}$, respectively. $\delta_n$ denotes the Kronecker delta function. $\binom{n}{k} = \frac{n!}{k!(n-k)!}$ denotes the combination number. $\mathcal{A} \cap \mathcal{B} = \varnothing $ means that the intersection of two sets $\mathcal{A}$ and $\mathcal{B}$ is the empty set.

\section{System Model}

\subsection{Synchronization Signal Model}

Consider initial downlink synchronization in a single cell, where the BS periodically transmits downlink synchronization signals, and a UT detects the presence of these synchronization signals in received signals to keep temporal synchronization with the BS. This is a typical arrangement in many existing cellular systems such as LTE \cite{S.Sesia2009}.

We call the time duration that the BS transmits synchronization signals each time as a synchronization time slot. As an example in LTE, a synchronization time slot corresponds to an orthogonal frequency division multiplexing (OFDM) symbol period \cite{S.Sesia2009}. The UT is assumed to utilize $K$ consecutive synchronization time slots to synchronize with the BS, where $K$ is an adjustable parameter yielding a tradeoff between the latency time and the success probability of initial synchronization.

At the $k$th synchronization time slot for $k=1,2,\ldots,K$, the discrete-time complex baseband signals can be modeled as
\begin{align}
{{\bf{Y}}_{k}}(\tau) = \left\{ {\begin{matrix*}[l]
{{\bf{F}}_k^H{{\bf{H}}_k}{{\bf{W}}_k}{{\bf{X}}_k} + {\bf{F}}_k^H{{\bf{Z}}_k},}&{ {\mathcal{H}_1: \tau = \tau_0}}\\
{{\bf{F}}_k^H{{\bf{Z}}_k},}&{ \mathcal{H}_0:\tau \neq \tau_0}
\end{matrix*}} \right. \label{Sync Signal Model}
\end{align}
where ${{\bf{X}}_k} \in {\mathbb{C}^{{N_{{\rm{t}}}} \times L}}$ denotes the synchronization signal transmitted by the BS at the $k$th time slot, $N_\mathrm{t}$ is the number of transmit streams, i.e., the number of transmit antenna ports, $L$ is the length of ${\bf X}_k$, ${{\bf{W}}_k} \in {\mathbb{C}^{{M_{{\rm{t}}}} \times {N_{{\rm{t}}}}}}$ denotes the precoding matrix, ${{\bf{H}}_k} \in {\mathbb{C}^{{M_{\rm{r}}} \times {M_{\rm{t}}}}}$ denotes the channel matrix, which is assumed to be frequency flat and temporally static in each time slot, but may vary across different time slots, ${M_{\rm{r}}}$ and ${M_{\rm{t}}}$ are the number of UT antennas and the number of BS antennas, respectively, ${{\bf{Z}}_k} \in {\mathbb{C}^{{M_{\rm{r}}} \times L}}$ denotes the additive white Gaussian noise (AWGN) matrix with i.i.d. $\mathcal{CN}( {0,\nu} )$ entries, ${{\bf{F}}_k} \in {\mathbb{C}^{{M_{\rm{r}}} \times {N_{\rm{r}}}}}$ denotes the combining matrix, $N_\mathrm{r}$ is the number of receive streams, i.e., the number of radio frequency (RF) chains at the UT, ${{\bf{Y}}_{k}}(\tau) \in {\mathbb{C}^{{N_{\rm{r}}} \times L}}$ denotes the signal observed by the UT at timing offset $\tau$ after combining. In addition, hypothesis ${\mathcal{H}_1}$ represents that timing offset $\tau$ is equal to a correct value $\tau_0$, hence the received signal is aligned with the transmitted synchronization signal, and hypothesis ${\mathcal{H}_0}$ means an incorrect $\tau$, therefore the synchronization signal is misaligned or absent \cite{S.M.Kay1998,D.W.Bliss2010}.

In the previous studies on mmWave massive MIMO synchronization, as well as the protocols of the existing cellular systems, the number of transmit streams for the synchronization signal ${\bf X}_k$ was set as $N_\mathrm{t} = 1$ \cite{S.Sesia2009,C.N.Barati2015,C.N.Barati2016,C.Liu2016}. Take LTE for example, where ${\bf X}_k$ is a Zadoff-Chu (ZC) sequence with $N_\mathrm{t} = 1$ and $L = 63$ used for the primary synchronization signal \cite{S.Sesia2009}. In this paper, the value of $N_\mathrm{t}$ is not confined to $1$, and the benefits by doing so will be explained in Section V. In addition, without loss of generality, we assume that ${\bf X}_k$ satisfies
\begin{align}
{{\bf{X}}_k}{\bf{X}}_k^H = \frac{L}{N_\mathrm{t}}{{\bf{I}}_{{N_{\rm{t}}}}} \label{Sync Signal Orthogonal}
\end{align}
for each $k$. Otherwise, we can always let ${\bf X}_k$ satisfy (\ref{Sync Signal Orthogonal}) by properly adjusting the precoding matrix ${\bf W}_k$.

{\color{black}{As a remark, it is known that mmWave communications should rely on directional transmitting/receiving beams. At the initial synchronization stage, both of the BS and the UT should find the correct beam directions for payload data transmission. In this paper, we assume that the entire initial synchronization procedure is divided into three steps. At the first step, when a UT is power on or moving to a new cell, it should keep downlink time/frequency synchronization with the BS by utilizing downlink synchronization signals transmitted from this BS. At this step, the target is reliable synchronization, e.g., minimizing the MD probability, in a fixed latency time. At the second step, after synchronized with the BS, the UT needs to find the correct receiving/transmitting beam directions for downlink/uplink transmission. At the last step, the UT transmits uplink synchronization signals with the previously obtained beam directions, and the BS finds the correct receiving/transmitting beam directions for uplink/downlink transmission. In this paper, we focus on the first step.}}

\subsection{Channel Model}

Consider the typical geometric channel model \cite{A.M.Sayeed2002,D.Tse2005,M.R.Akdeniz2014}. Both of the BS and UT are assumed to be equipped with a uniform linear array (ULA). Then the channel matrix ${\bf H}_k$ in (\ref{Sync Signal Model}) can be expressed as
\begin{align}
{{\bf{H}}_k} = \sum\limits_{p = 1}^P {{\alpha _{p,k}}{\bf{u}}( {{\theta _{\mathrm{r},p}}} ){{{{\bf{v}}^H( {{\theta _{\mathrm{t},p}}} )} }}} \label{Channel Model 1}
\end{align}
where $P$ denotes the total number of paths, ${\alpha _{p,k}}$ denotes the complex fading gain of the $p$th path at the $k$th synchronization time slot, ${{\theta _{\mathrm{r},p}}}$ and ${{\theta _{\mathrm{t},p}}}$ denote the arrival and departure angles\footnote{Here $\theta$ represents the virtual angle, also known as spatial frequency. The relation between the virtual angle $\theta$ and the physical angle $\vartheta$ is $\theta = d \sin \vartheta / \lambda$, where $\lambda$ denotes the carrier wavelength and $d$ denotes the antenna space of the ULA. A typical value of $d$ is $d = \lambda/2$.} of the $p$th path, respectively, and the array response vectors are expressed as
\begin{align}
{\bf{u}}({\theta _{\rm{r}}}) &= {[ {1,{e^{j2\pi {\theta _{\rm{r}}}}}, \ldots ,{e^{j2\pi ({M_{\rm{r}}} - 1){\theta _{\rm{r}}}}}} ]^T}, \;\; 0 \leq \theta_\mathrm{r} \leq 1 \label{Channel Model 2} \\
{\bf{v}}({\theta _{\rm{t}}}) &= {[ {1,{e^{j2\pi {\theta _{\rm{t}}}}}, \ldots ,{e^{j2\pi ({M_{\rm{t}}} - 1){\theta _{\rm{t}}}}}} ]^T}, \;\; 0 \leq \theta_\mathrm{t} \leq 1. \label{Channel Model 3}
\end{align}
The complex fading gain $\alpha_{p,k}$ of the $p$th path is assumed to follow complex Gaussian distribution with zero mean and variance $\beta _p$, i.e.,
\begin{align}
{\alpha _{p,k}} \sim \mathcal{CN}( {0,{\beta _p}} ). \label{Channel Model 4}
\end{align}
The arrival and departure angles ${{\theta _{\mathrm{r},p}}}$ and ${{\theta _{\mathrm{t},p}}}$ of each path are assumed to keep constant within these $K$ synchronization time slots since they usually vary slowly compared with the fast fading \cite{Y.Wang2015,I.Viering2002,M.Nicoli2003}. Different paths are assumed to be uncorrelated, and the temporal correlation of each path at the $k$th and $l$th synchronization time slots is described as $\psi_{k,l}$, i.e.,
\begin{align}
\mathbb{E}\{ {{\alpha _{p,k}}\alpha _{q,l}^*} \} = \left\{ {\begin{matrix*}[l]
{{\beta _p}{\psi _{k,l}},}&{{\text{if }}p = q}\\
{0,}&{\text{otherwise},}
\end{matrix*}} \right. \label{Channel Model 5}
\end{align}
where
\begin{align}
\psi_{k,k} = 1 \label{Channel Model 6}
\end{align}
and $|\psi_{k,l}|\leq 1$ if $k \neq l$. In addition, the total average gain of all the $P$ paths is assumed to be normalized, i.e.,
\begin{align}
\sum\limits_{p = 1}^P {{\beta _p}}  = 1 . \label{Channel Model 7}
\end{align}

\section{Basic Requirements of Precoding and Combining Matrices}

\subsection{Implementation Architecture Constraint}

In mmWave massive MIMO systems, the precoding and combining are usually implemented in hybrid analog-digital architectures \cite{A.Alkhateeb2014,S.Han2015,O.E.Ayach2014}. Taking precoding for example, the precoding matrix in (\ref{Sync Signal Model}) can be decomposed as ${\bf W}_k = {\bf W}_{k,\mathrm{RF}} {\bf W}_{k,\mathrm{BB}} $, where ${\bf W}_{k,\mathrm{RF}} \in \mathbb{C}^{M_\mathrm{t} \times N_{\rm RF}}$ and ${\bf W}_{k,\mathrm{BB}} \in \mathbb{C}^{N_{\mathrm{RF}} \times N_{\rm t}}$ represent the RF and baseband precoding matrices, implemented in the analog and digital domains, respectively, and $N_{\rm RF}$ is the number of RF chains satisfying $M_{\rm t} \leq N_{\rm{RF}} \leq N_{\rm t}$. In this paper, we assume that both of the precoding and combining are implemented in the analog domain using networks of phase shifters. Therefore all the entries in the precoding and combining matrices should have constant amplitude, i.e.,
\begin{align}
{\big| {{{[ {{{\bf{W}}_k}} ]_{m,n}}}} \big|} &= \frac{1}{\sqrt{{{M_{\rm{t}}}}}} ,\;\;  \forall \; m,n,k \label{Constant Amplitude Constraint Precoding Matrix} \\
{\big| {{{[ {{{\bf{F}}_k}} ]_{m,n}}}} \big|} &= \frac{1}{\sqrt{{{M_{\rm{r}}}}}} ,\;\;  \forall \; m,n,k. \label{Constant Amplitude Constraint Combining Matrix}
\end{align}
Note that this analog-only architecture is a more strict constraint than hybrid architectures, and the designed precoding and combining matrices under this architecture can also be easily incorporated in hybrid architectures. Still taking precoding for example, once ${\bf W}_k $ has been designed satisfying (\ref{Constant Amplitude Constraint Precoding Matrix}), we can let ${\bf W}_{k,\rm{RF}} = [{\bf W}_k,{\bf 0}_{M_{\rm t}\times({N_{\rm{RF}}-N_{\rm t}})}]$ and ${\bf W}_{k,\rm{BB}} = [{\bf I}_{N_{\rm{t}}},{\bf 0}_{N_{\rm t}\times(N_{\rm {RF}}-N_{\rm t})}]^T$, i.e., selecting $N_{\rm t}$ RF chains from the total $N_{\rm {RF}}$ RF chains to implement ${\bf W}_k $ in hybrid architectures.

\subsection{Omnidirectional Coverage over Total $K$ Time Slots}

To overcome the high isotropic path loss in mmWave frequency bands and extend the transmission range, mmWave massive MIMO systems usually rely on highly directional transmission. {\color{black}{However, in the initial synchronization stage, neither of the BS and UT knows which departure or arrival direction should be preferred for transmitting or receiving. Therefore, both of them should transmit and receive the synchronization signals omnidirectionally to guarantee reliable coverage.}}

Assume that there is only one single path in (\ref{Channel Model 1}) with departure angle $\theta_\mathrm{t}$ and arrival angle $\theta_\mathrm{r}$, i.e., ${{\bf{H}}_k} = {\alpha _k}{\bf{u}}({\theta _{\rm{r}}}){{\bf{v}}^H}({\theta _{\rm{t}}})$. Recall (\ref{Sync Signal Model}) and under $\mathcal{H}_1$, the received signal at the UT side at the $k$th synchronization time slot without AWGN is ${\bf{F}}_k^H{{\bf{H}}_k}{{\bf{W}}_k}{{\bf{X}}_k} = {\alpha _k}{\bf{F}}_k^H{\bf{u}}({\theta _{\rm{r}}}){{\bf{v}}^H}({\theta _{\rm{t}}}){{\bf{W}}_k}{{\bf{X}}_k}$. Then, the sum average signal power over the total $K$ synchronization time slots is
\begin{align}
P &= \sum\limits_{k = 1}^K {{P_k}}  = \sum\limits_{k = 1}^K {\mathbb{E}\{ \|{{\alpha _k}{\bf{F}}_k^H{\bf{u}}({\theta _{\rm{r}}}){{\bf{v}}^H}({\theta _{\rm{t}}}){{\bf{W}}_k}{{\bf{X}}_k}} \|_{\rm{F}}^2\}} \nonumber \\
&= \frac{L}{{{N_{\rm{t}}}}}\sum\limits_{k = 1}^K { \mathbb{E}\{|{\alpha _k}|^2\}{{\bf{v}}^H}({\theta _{\rm{t}}}){{\bf{W}}_k}{\bf{W}}_k^H{\bf{v}}({\theta _{\rm{t}}}){{\bf{u}}^H}({\theta _{\rm{r}}}){{\bf{F}}_k}{\bf{F}}_k^H{\bf{u}}({\theta _{\rm{r}}})} \nonumber \\
&= \frac{L}{{{N_{\rm{t}}}}}\sum\limits_{k = 1}^K {{{\bf{v}}^H}({\theta _{\rm{t}}}){{\bf{W}}_k}{\bf{W}}_k^H{\bf{v}}({\theta _{\rm{t}}}){{\bf{u}}^H}({\theta _{\rm{r}}}){{\bf{F}}_k}{\bf{F}}_k^H{\bf{u}}({\theta _{\rm{r}}})} \label{Sum Average Power}
\end{align}
where the average is taken over $\alpha_k$, the third equality is from (\ref{Sync Signal Orthogonal}), and the last equality is from (\ref{Channel Model 5}) and (\ref{Channel Model 7}). It is expected that the sum average power (\ref{Sum Average Power}) is constant for any departure angle $\theta_\mathrm{t}$ and arrival angle $\theta_\mathrm{r}$, i.e.,
\begin{align}
\sum\limits_{k = 1}^K {{{\bf{v}}^H}({\theta _{\rm{t}}}){{\bf{W}}_k}{\bf{W}}_k^H{\bf{v}}({\theta _{\rm{t}}})\cdot{{\bf{u}}^H}({\theta _{\rm{r}}}){{\bf{F}}_k}{\bf{F}}_k^H{\bf{u}}({\theta _{\rm{r}}})}  = c, \;\; \forall \; {\theta _{\rm{r}}},{\theta _{\rm{t}}} \in [0,1]. \label{Average Omnidirectional Constraint 1}
\end{align}
To determine the unknown coefficient $c$ therein, we integrate over ${\theta _{\rm{r}}}$ and ${\theta _{\rm{t}}}$ at both sides of (\ref{Average Omnidirectional Constraint 1}). The right hand side is $\int_0^1 {\int_0^1 c \mathrm{d}{\theta _{\rm{r}}}\mathrm{d}{\theta _{\rm{t}}}}  = c$, and the left hand side can be expressed as
\begin{align}
&\sum\limits_{k = 1}^K {\int_0^1 {{{{\bf{u}}^H({\theta _{\rm{r}}})}}{{\bf{F}}_k}{\bf{F}}_k^H{\bf{u}}({\theta _{\rm{r}}})\mathrm{d}{\theta _{\rm{r}}}} \int_0^1 {{{{\bf{v}}^H({\theta _{\rm{t}}})}}{{\bf{W}}_k}{\bf{W}}_k^H{\bf{v}}({\theta _{\rm{t}}})\mathrm{d}{\theta _{\rm{t}}}} } \nonumber \\
&= \sum\limits_{k = 1}^K {\mathrm{tr}\bigg( {{{\bf{F}}_k}{\bf{F}}_k^H {\int_0^1 {{\bf{u}}({\theta _{\rm{r}}}){{{\bf{u}}^H({\theta _{\rm{r}}})}}\mathrm{d}{\theta _{\rm{r}}}} } } \bigg)\mathrm{tr}\bigg( {{{\bf{W}}_k}{\bf{W}}_k^H {\int_0^1 {{\bf{v}}({\theta _{\rm{t}}}){{{\bf{v}}^H({\theta _{\rm{t}}})}}\mathrm{d}{\theta _{\rm{t}}}} } } \bigg)} \nonumber \\
&= \sum\limits_{k = 1}^K {\mathrm{tr}( {{\bf{F}}_k}{{\bf{F}}_k^H} )\mathrm{tr}( {{\bf{W}}_k}{{\bf{W}}_k^H} )} = K N_\mathrm{r} N_\mathrm{t} \nonumber
\end{align}
where the first equality is from the fact that $\mathrm{tr}({\bf{AB}}) = \mathrm{tr}({\bf{BA}})$, the second equality is because ${\int_0^1 {{\bf{u}}({\theta _{\rm{r}}}){{{\bf{u}}^H({\theta _{\rm{r}}})}}\mathrm{d}{\theta _{\rm{r}}}} } = {\bf I}_{M_\mathrm{r}}$ and ${\int_0^1 {{\bf{v}}({\theta _{\rm{t}}}){{{\bf{v}}^H({\theta _{\rm{t}}})}}\mathrm{d}{\theta _{\rm{t}}}} } = {\bf I}_{M_\mathrm{t}}$, which can be obtained from (\ref{Channel Model 2}) and (\ref{Channel Model 3}) immediately, and the last equality can be obtained from (\ref{Constant Amplitude Constraint Precoding Matrix}) and (\ref{Constant Amplitude Constraint Combining Matrix}). Therefore, we have $c = K N_\mathrm{r} N_\mathrm{t}$ and can rewrite (\ref{Average Omnidirectional Constraint 1}) as
\begin{align}
\sum\limits_{k = 1}^K {{{{\bf{u}}^H({\theta _{\rm{r}}})}}{{\bf{F}}_k}{\bf{F}}_k^H{\bf{u}}({\theta _{\rm{r}}}) \cdot {{{\bf{v}}^H({\theta _{\rm{t}}})}}{{\bf{W}}_k}{\bf{W}}_k^H{\bf{v}}({\theta _{\rm{t}}})}  = K N_\mathrm{r} N_\mathrm{t}, \;\; \forall \; {\theta _{\rm{r}}},{\theta _{\rm{t}}} \in [0,1]. \label{Average Omnidirectional Constraint 2}
\end{align}

\begin{figure}[htbp]
\centering
\includegraphics[width=0.5\columnwidth]{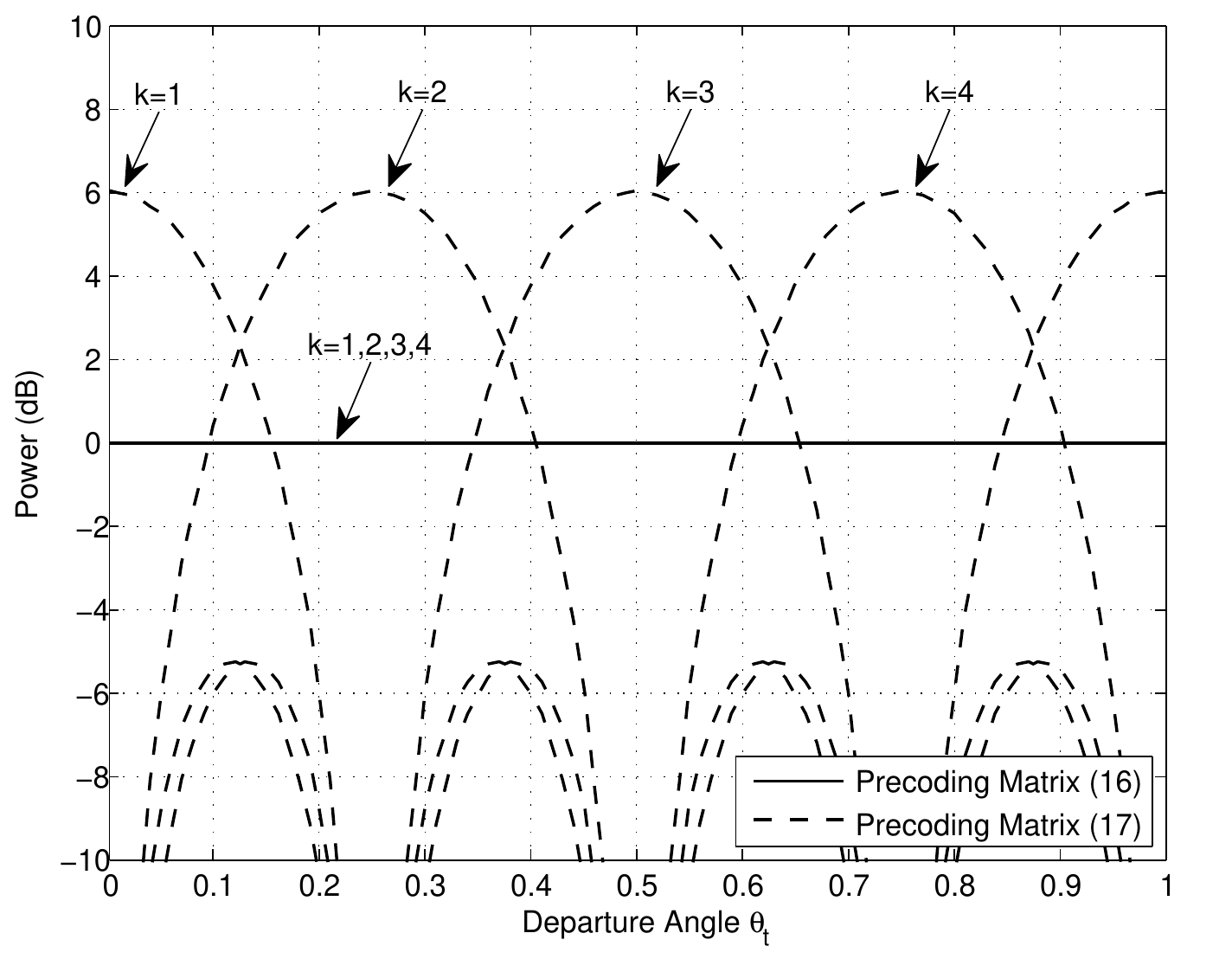}
\caption{Comparison of transmit power between two different precoding matrices.}
\end{figure}

Note that condition (\ref{Average Omnidirectional Constraint 2}) implies that the omnidirectional coverage is guaranteed in total $K$ synchronization time slots in an average sense. Hence we call it average omnidirectional coverage, and it does not necessarily require wide beams at both of the BS and UT sides when $K > 1$, since sweeping a narrow beam towards different directions at different time slots can also achieve average omnidirectional coverage. For example, in the case with $M_\mathrm{t} = K = 4$ and $M_\mathrm{r} = N_\mathrm{r} = N_\mathrm{t} = 1$, since there is only one single receive antenna, only precoding needs to be considered. Then (\ref{Average Omnidirectional Constraint 2}) becomes
\begin{align}
\sum\limits_{k = 1}^K {{{{\bf{u}}^H({\theta _{\rm{r}}})}}{{\bf{F}}_k}{\bf{F}}_k^H{\bf{u}}({\theta _{\rm{r}}})  }  = K N_\mathrm{t}, \;\; \forall \; {\theta _{\rm{t}}} \in [0,1]. \label{Average Omnidirectional Constraint 3}
\end{align}
For the following two precoding matrices
\begin{align}
{\bf{W}}_k^{(1)} &= {[{{\bf{I}}_4}]_{:,k}} \label{Precoding Matrix Example 1} \\
{\bf{W}}_k^{(2)} &= \frac{1}{2}  {[1,{e^{j\pi k/2}},{e^{j\pi k}},{e^{j3\pi k/2}}]^T} , \label{Precoding Matrix Example 2}
\end{align}
the corresponding transmit power at each departure angle $\theta_{\rm t}$ can be obtained according to ${{{\bf{v}}^H}({\theta _{\rm{t}}}){{\bf{W}}_k}{\bf{W}}_k^H{\bf{v}}({\theta _{\rm{t}}})}$ for each time slot $k = 1,2,3,4$, and are plotted in Fig. 1. It can be observed that, both precoding matrices (\ref{Precoding Matrix Example 1}) and (\ref{Precoding Matrix Example 2}) can guarantee average omnidirectional coverage constraint (\ref{Average Omnidirectional Constraint 3}) over total $4$ synchronization time slots. The difference is that, (\ref{Precoding Matrix Example 1}) guarantees omnidirectional coverage in every synchronization time slot, while (\ref{Precoding Matrix Example 2}) generates narrow beams toward different directions in different synchronization time slots.

\section{Effect of Precoding and Combining Matrices to Synchronization Performance}

\subsection{GLRT Based Synchronization Detector}

As described in (\ref{Sync Signal Model}), we model the temporal synchronization as a binary hypothesis test problem \cite{S.M.Kay1998,D.W.Bliss2010}. Since there are unknown parameters in (\ref{Sync Signal Model}), including the noise variance ${{\nu}}$ and the effective channel matrix ${{\bf{G}}_k} \triangleq {\bf{F}}_k^H{{\bf{H}}_k}{{\bf{W}}_k}$, we use GLRT to perform synchronization detection \cite{S.M.Kay1998,D.W.Bliss2010}. The test statistic under GLRT is defined as
\begin{align}
T'(\tau) = \ln \frac{{\mathop {\max }\nolimits_{{\bf{G}},{\nu}} f( {{\bf{Y}}(\tau)|{\mathcal{H}_1},{\bf{G}},{\nu}} )}}{{\mathop {\max }\nolimits_{{\nu}} f( {{\bf{Y}}(\tau)|{\mathcal{H}_0},{\nu}} )}} \overset{\mathcal{H}_1}{\underset{\mathcal{H}_0}{\gtrless}}  \gamma ' \label{GLRT Detector Raw}
\end{align}
where ${\bf{Y}}(\tau) = [ {{{\bf{Y}}_{1}}(\tau),{{\bf{Y}}_{2}}(\tau), \ldots ,{{\bf{Y}}_{K}}(\tau)} ]$, ${\bf{G}} = [ {{{\bf{G}}_1},{{\bf{G}}_2}, \ldots ,{{\bf{G}}_K}} ]$, $\gamma'$ is a threshold value. Then we have the following theorem.

\emph{Theorem 1:} The test statistic (\ref{GLRT Detector Raw}) is equivalent to
\begin{align}
T(\tau) = \frac{{\sum\nolimits_{k = 1}^K {\mathrm{tr}( {{{\bf{Y}}_{k}(\tau)}{\bf{X}}_k^H{{( {{{\bf{X}}_k}{\bf{X}}_k^H} )^{ - 1}}}{{\bf{X}}_k}{\bf{Y}}_{k}^H(\tau){{( {{\bf{F}}_k^H{{\bf{F}}_k}} )^{ - 1}}}} )} }}{{\sum\nolimits_{k = 1}^K {\mathrm{tr}( {{{\bf{Y}}_{k}(\tau)}{\bf{Y}}_{k}^H(\tau){{( {{\bf{F}}_k^H{{\bf{F}}_k}} )^{ - 1}}}} )} }} \overset{\mathcal{H}_1}{\underset{\mathcal{H}_0}{\gtrless}}  \gamma \label{GLRT Detector}
\end{align}
where $\gamma = 1 - \exp \big( { - \frac{{\gamma '}}{{KL{N_{\rm{r}}}}}} \big)$.
\begin{IEEEproof}
See Appendix A.
\end{IEEEproof}

According to (\ref{GLRT Detector}), the synchronization detector at the UT is operated as follows. The observed signal ${\bf Y}(\tau)$ under timing offset $\tau$ is used to evaluate the test statistic $T(\tau)$, which is then compared with the threshold value $\gamma$. If $T(\tau)$ is greater than $\gamma$, successful synchronization can be claimed. Otherwise, the UT adjusts its timing offset $\tau$, e.g., letting $\tau = \tau+1$, and re-execute the above procedure until a successful synchronization appears.

The performance of  temporal synchronization is typically characterized by the probability of MD given the correct timing offset, versus the probability of FA that occurs if synchronization is declared in error. The MD probability and the FA probability are, respectively, defined as
\begin{align}
{P_{{\rm{MD}}}} &= \mathbb{P}\{ {T(\tau) < \gamma |{\mathcal{H}_1}} \} \label{MD Probability Definition} \\
{P_{{\rm{FA}}}} &= \mathbb{P}\{ {T(\tau) > \gamma |{\mathcal{H}_0}} \}. \label{FA Probability Definition}
\end{align}
In the following subsections, we will investigate the effect of the precoding and combining matrices to the MD and FA probabilities.

\subsection{MD Probability}

In this subsection, we mainly investigate the effect of the recoding and combining matrices to the MD probability. First, we present the following theorem.

\emph{Theorem 2:} The MD probability (\ref{MD Probability Definition}) can be expressed as
\begin{align}
P_{\mathrm{MD}} = \mathbb{P}\Bigg\{ {\frac{{\sum\nolimits_{k = 1}^K {\| {{{(L/N_{\rm t})^{1/2}}}{\bf{F}}_k^H{{\bf{H}}_k}{{\bf{W}}_k} + {{\bf{Z}}_{k,2}}} \|_{\rm{F}}^2} }}{{\sum\nolimits_{k = 1}^K {\| {{{\bf{Z}}_{k,1}}} \|_{\rm{F}}^2} }} < \frac{\gamma}{1-\gamma} } \Bigg\} \label{MD Probability}
\end{align}
where ${\bf Z}_{k,1} \in \mathbb{C}^{N_\mathrm{r} \times (L-N_\mathrm{t})}$ and ${\bf Z}_{k,2} \in \mathbb{C}^{N_\mathrm{r} \times N_\mathrm{t}}$ are both with i.i.d. $\mathcal{CN}(0,\nu)$ entries and they are independent of each other.
\begin{IEEEproof}
See Appendix B.
\end{IEEEproof}

By letting
\begin{align}
{\bf{g}} &= \mathrm{vec}\{ {[ {{\bf{ F}}_1^H{{\bf{H}}_1}{{\bf{W}}_1}, \ldots ,{\bf{ F}}_K^H{{\bf{H}}_K}{{\bf{W}}_K}} ]} \} \label{Effective Channel Vector} \\
{{\bf{z}}_1} &= \mathrm{vec}\{ {[ {{{\bf{Z}}_{1,1}},{{\bf{Z}}_{2,1}}, \ldots ,{{\bf{Z}}_{K,1}}} ]} \} \nonumber \\
{{\bf{z}}_2} &= \mathrm{vec}\{ {[ {{{\bf{Z}}_{1,2}},{{\bf{Z}}_{2,2}}, \ldots ,{{\bf{Z}}_{K,2}}} ]} \}, \nonumber
\end{align}
we can express (\ref{MD Probability}) as
\begin{align}
P_{\mathrm{MD}} = \mathbb{P}\Bigg\{ {\frac{{{{( { {(L/N_{\rm t})^{1/2}} {\bf{g}} + {{\bf{z}}_2}} )^H}}( { {(L/N_{\rm t})^{1/2}} {\bf{g}} + {{\bf{z}}_2}} )}}{{{\bf{z}}_1^H{{\bf{z}}_1}}} < \frac{\gamma}{1-\gamma} } \Bigg\} \label{MD Probability Simple}
\end{align}
where ${\bf z}_1 \sim \mathcal{CN}({\bf 0},\nu{\bf I}_{KN_\mathrm{r}(L-N_\mathrm{t})})$ and ${\bf z}_2 \sim \mathcal{CN}({\bf 0},\nu{\bf I}_{KN_\mathrm{r}N_\mathrm{t}})$. According to (\ref{Channel Model 1}), (\ref{Channel Model 4}), and (\ref{Effective Channel Vector}), we have ${\bf g} \sim \mathcal{CN}({\bf 0},{\bf R})$ where the covariance matrix can be expressed as
\begin{align}
{\bf{R}} = \mathbb{E}\{ {{\bf{g}}{{\bf{g}}^H}} \} = \left[ {\begin{matrix}
{{{\bf{R}}_{1,1}}}&{{{\bf{R}}_{1,2}}}& \cdots &{{{\bf{R}}_{1,K}}}\\
{{{\bf{R}}_{2,1}}}&{{{\bf{R}}_{2,2}}}& \cdots &{{{\bf{R}}_{2,K}}}\\
 \vdots & \vdots & \ddots & \vdots \\
{{{\bf{R}}_{K,1}}}&{{{\bf{R}}_{K,2}}}& \cdots &{{{\bf{R}}_{K,K}}}
\end{matrix}} \right] . \label{Effective Channel Vector Covariance}
\end{align}
The $(k,l)$th subblock in (\ref{Effective Channel Vector Covariance}) is defined as
\begin{align}
{{\bf{R}}_{k,l}} &= \mathbb{E}\{ {\mathrm{vec}( {{\bf{ F}}_k^H{{\bf{H}}_k}{{\bf{W}}_k}} ){{( {\mathrm{vec}( {{\bf{ F}}_l^H{{\bf{H}}_l}{{\bf{W}}_l}} )} )^H}}} \} \nonumber \\
&= ( {{\bf{W}}_k^T \otimes {\bf{ F}}_k^H} )\mathbb{E}\{ {\mathrm{vec}( {{{\bf{H}}_k}} ){{( {\mathrm{vec}( {{{\bf{H}}_l}} )} )^H}}} \}{( {{\bf{W}}_l^* \otimes {\bf{ F}}_l} )} \label{Effective Channel Vector Covariance Subblock}
\end{align}
where the last equality is with the fact that $\mathrm{vec}( {{\bf{ABC}}} ) = ( {{{\bf{C}}^T} \otimes {\bf{A}}} )\mathrm{vec}( {\bf{B}} )$. The middle term in (\ref{Effective Channel Vector Covariance Subblock}) can be further expressed as
\begin{align}
\mathbb{E}\{ {\mathrm{vec}( {{{\bf{H}}_k}} ){{( {\mathrm{vec}( {{{\bf{H}}_l}} )} )^H}}} \} &= \mathbb{E}\Bigg\{ {\Bigg( {\sum\limits_{p = 1}^P {{\alpha _{p,k}}{{{\bf{v}}^*({\theta _{{\rm{t}},p}})}} \otimes {\bf{u}}({\theta _{{\rm{r}},p}})} } \Bigg){{\Bigg( {\sum\limits_{p = 1}^P {{\alpha _{p,k}}{{{\bf{v}}^*({\theta _{{\rm{t}},p}})}} \otimes {\bf{u}}({\theta _{{\rm{r}},p}})} } \Bigg)^H}}} \Bigg\} \nonumber \\
&= {\psi _{k,l}}\sum\limits_{p = 1}^P {{\beta _p}( {{{{\bf{v}}^*({\theta _{{\rm{t}},p}})}} \otimes {\bf{u}}({\theta _{{\rm{r}},p}})} )( {{{{\bf{v}}^T({\theta _{{\rm{t}},p}})}} \otimes {{{\bf{u}}^H({\theta _{{\rm{r}},p}})}}} )} \label{Channel Vector Covariance}
\end{align}
where the first equality is from (\ref{Channel Model 1}) and the last equality is from (\ref{Channel Model 5}).

From (\ref{MD Probability Simple}), (\ref{Effective Channel Vector Covariance}), and (\ref{Effective Channel Vector Covariance Subblock}), it can be observed that the precoding and combining matrices over total $K$ synchronization time slots $\{ {\bf W}_k , {\bf F}_k \}_{k=1}^K$ mainly affect the covariance matrix $\bf R$ of the effective channel vector $\bf g$, thereby affecting the MD probability $P_{\mathrm{MD}}$. To further quantify this effect, we need to derive an exact expression for (\ref{MD Probability Simple}). Before proceeding, we present the following useful lemma.

\emph{Lemma 1:} Consider two random variables $X = \sum\nolimits_{m=1}^{M} |x_m|^2$ and $Y = \sum\nolimits_{n=1}^{N} |y_n|^2$ with independently distributed entries $x_m \sim \mathcal{CN}(0,\lambda_m)$ and $y_n \sim \mathcal{CN}(0,\sigma_n)$. Then the probability $\mathbb{P}\{ {\frac{X}{Y} < t} \}$ can be approximated by
\begin{align}
\mathbb{P}\bigg\{ {\frac{X}{Y} < t} \bigg\} \approx {t^{M}} {\prod\limits_{m = 1}^{M} {\frac{1}{{{\lambda _m}}}} } \sum\limits_{{l_1} + {l_2} +  \cdots  + {l_N} = M} { \prod \limits_{n=1}^{N} \sigma _n^{{l_n}}} \label{Generalized F Distribution CDF Asymptotic}
\end{align}
when $t$ is small.
\begin{IEEEproof}
See Appendix C.
\end{IEEEproof}
Note that the $M$ components $x_1,x_2,\ldots,x_M$ in $X$, as well as the $N$ components $y_1,y_2,\ldots,y_N$ in $Y$, may have different variances. Therefore, $X$ and $Y$ do not follow the standard Chi-squared distribution. Hence, $X/Y$ does not follow the traditional F-distribution. The above lemma is then used to establish the following theorem.

\emph{Theorem 3:} Let $r$ and $\{\lambda_m \}_{m=1}^{r}$ denote the rank and the non-zero eigenvalues of $\bf R$, respectively. Then the MD probability (\ref{MD Probability Simple}) can be asymptotically expressed as
\begin{align}
P_{\mathrm{MD}} \approx {   \bigg( \frac{{N_{\rm t} \nu \gamma } }{L(1-\gamma)} \bigg)^r} \binom{{KL{N_{\rm{r}}} - 1}}{r} \prod\limits_{m = 1}^r \lambda_m^{-1} \label{MD Probability Asymptotic}
\end{align}
{\color{black}{when $\gamma$ is small,}} where $\binom{n}{k} = \frac{n!}{k!(n-k)!}$ denotes the combination number.
\begin{IEEEproof}
It can be obtained from Lemma 1 immediately.
\end{IEEEproof}

The asymptotic result (\ref{MD Probability Asymptotic}) in Theorem 1 presents a simple and useful tool to analyze the effect of the precoding and combining matrices to the MD probability. From (\ref{MD Probability Asymptotic}), it can be observed that $P_{\mathrm{MD}}$ is mainly affected by the rank $r$ and the non-zero eigenvalues $\{\lambda_m \}_{m=1}^{r}$ of $\bf R$. If we regard $P_{\mathrm{MD}}$ as a function of the signal-to-noise ratio (SNR) value $1/\nu$, the slope of the curve in the high-SNR regime as in (\ref{MD Probability Asymptotic}) will be determined by $r$, and the horizontal shift of the curve will be determined by $\prod\nolimits_{m = 1}^r {\lambda _m^{ - 1}} $. This is similar to the concepts of diversity and coding gains of traditional space-time block codes (STBCs), where a good STBC design should first maximize the diversity gain, and then maximize the coding gain \cite{V.Tarokh1998}. Here, the same with the design criteria for STBC, to minimize the asymptotic $P_{\mathrm{MD}}$ (\ref{MD Probability Asymptotic}), it is required that the rank $r$ of $\bf R$ should be maximized first, and then the product of the non-zero eigenvalues $\prod\nolimits_{m = 1}^r {\lambda _m} $ of $\bf R$ should be maximized.

However, from (\ref{Effective Channel Vector Covariance}) and (\ref{Effective Channel Vector Covariance Subblock}) one can see that the rank $r$ and the non-zero eigenvalues $\{\lambda_m \}_{m=1}^{r}$ of $\bf R$ are determined not only by the precoding and combining matrices $\{ {{\bf{W}}_k},{{\bf{F}}_k}\} _{k = 1}^K$, but also by the channel covariance matrix $\mathbb{E}\{ \mathrm{vec}({{\bf{H}}_k}){(\mathrm{vec}({{\bf{H}}_l}))^H}\} $. This covariance matrix usually depends on the transmission scenario and the characteristics of the terrain, and cannot be foreknown in the initial synchronization stage. Therefore, we consider two typical models, including the single-path channel and the i.i.d. channel. These two models represent two extreme transmission scenarios, where the former is with a single sparse path while the latter is with sufficiently rich paths. Note that these two models are mainly used to simplify the theory analysis, and in numerical simulations we will also use more realistic channel models.

\subsubsection{The single-path channel}

In this case, there is only $P=1$ path in (\ref{Channel Model 1}), and then (\ref{Channel Vector Covariance}) becomes
\begin{align}
\mathbb{E}\{ {\mathrm{vec}( {{{\bf{H}}_k}} ){{( {\mathrm{vec}( {{{\bf{H}}_l}} )} )^H}}} \} = {\psi _{k,l}}({{\bf{v}}^*}({\theta _{\rm{t}}}) \otimes {\bf{u}}({\theta _{\rm{r}}}))({{\bf{v}}^T}({\theta _{\rm{t}}}) \otimes {{\bf{u}}^H}({\theta _{\rm{r}}})) \label{Channel Vector Covariance Single Path}
\end{align}
where we have utilized (\ref{Channel Model 7}) and omitted the path index $p$ for notational simplicity. Substituting (\ref{Channel Vector Covariance Single Path}) into (\ref{Effective Channel Vector Covariance Subblock}) yields
\begin{align}
{{\bf{R}}_{k,l}} = {\psi _{k,l}}(\underbrace {{\bf{W}}_k^T{{\bf{v}}^*}({\theta _{\rm{t}}}) \otimes {\bf{F}}_k^H{\bf{u}}({\theta _{\rm{r}}})}_{{{\bf{a}}_k}})(\underbrace {{{\bf{v}}^T}({\theta _{\rm{t}}}){\bf{W}}_l^* \otimes {{\bf{u}}^H}({\theta _{\rm{r}}}){{\bf{F}}_l}}_{{\bf{a}}_l^H}). \label{Effective Channel Vector Covariance Subblock Single Path}
\end{align}
Let
\begin{align}
{\bf{A}} = \left[ {\begin{matrix}
{{{\bf{a}}_1}}&{\bf{0}}& \cdots &{\bf{0}}\\
{\bf{0}}&{{{\bf{a}}_2}}& \cdots &{\bf{0}}\\
 \vdots & \vdots & \ddots & \vdots \\
{\bf{0}}&{\bf{0}}& \cdots &{{{\bf{a}}_K}}
\end{matrix}} \right] \nonumber
\end{align}
and $[{\bf \Psi}]_{k,l} = \psi_{k,l}$.
Then with (\ref{Effective Channel Vector Covariance Subblock Single Path}), we can express (\ref{Effective Channel Vector Covariance}) as
\begin{align}
{\bf R} = {\bf A}{\bf \Psi}{\bf A}^H . \label{Effective Channel Vector Covariance Single Path}
\end{align}
Assume ${\bf \Psi}$ is of full rank $K$. According to the property that two matrices $\bf XY$ and $\bf YX$ have the same rank and non-zero eigenvalues when ${\bf X}$ is invertible, we know that the rank and the non-zero eigenvalues of $\bf R$ in (\ref{Effective Channel Vector Covariance Single Path}) are identical to those of
\begin{align}
\tilde{\bf{ R}} = {\bf \Psi}{\bf A}^H{\bf A} = {\bf \Psi} \cdot {\rm diag} \{ {\bf{a}}_1^H{{\bf{a}}_1},{\bf{a}}_2^H{{\bf{a}}_2},\ldots,{\bf{a}}_K^H{{\bf{a}}_K}\} . \label{Effective Channel Vector Covariance Single Path Equivalent}
\end{align}
With (\ref{Effective Channel Vector Covariance Subblock Single Path}) we know that
\begin{align}
{\bf{a}}_k^H{{\bf{a}}_k} &= ({{\bf{v}}^T}({\theta _{\rm{t}}}){\bf{W}}_k^* \otimes {{\bf{u}}^H}({\theta _{\rm{r}}}){{\bf{F}}_k})({\bf{W}}_k^T{{\bf{v}}^*}({\theta _{\rm{t}}}) \otimes {\bf{F}}_k^H{\bf{u}}({\theta _{\rm{r}}})) \nonumber \\
&= {{\bf{v}}^T}({\theta _{\rm{t}}}){\bf{W}}_k^*{\bf{W}}_k^T{{\bf{v}}^*}({\theta _{\rm{t}}}) \otimes {{\bf{u}}^H}({\theta _{\rm{r}}}){{\bf{F}}_k}{\bf{F}}_k^H{\bf{u}}({\theta _{\rm{r}}}) \nonumber \\
&= {{\bf{v}}^H}({\theta _{\rm{t}}}){{\bf{W}}_k}{\bf{W}}_k^H{\bf{v}}({\theta _{\rm{t}}}) \cdot {{\bf{u}}^H}({\theta _{\rm{r}}}){{\bf{F}}_k}{\bf{F}}_k^H{\bf{u}}({\theta _{\rm{r}}}) , \label{Transceive Power Gain}
\end{align}
where the second equality is because $({\bf{A}} \otimes {\bf{B}})({\bf{C}} \otimes {\bf{D}}) = {\bf{AC}} \otimes {\bf{BD}}$, and the last equality is because $a \otimes b = ab$ for two scalars $a$ and $b$, and ${{\bf{v}}^T}({\theta _{\rm{t}}}){\bf{W}}_k^*{\bf{W}}_k^T{{\bf{v}}^*}({\theta _{\rm{t}}})$ is equal to its transpose ${{\bf{v}}^H}({\theta _{\rm{t}}}){{\bf{W}}_k}{\bf{W}}_k^H{\bf{v}}({\theta _{\rm{t}}})$. With (\ref{Transceive Power Gain}) and recall (\ref{Average Omnidirectional Constraint 2}), it should be satisfied that
\begin{align}
\sum\limits_{k = 1}^K {{\bf{a}}_k^H{{\bf{a}}_k}}  = KN_{\rm r}N_{\rm t}. \label{Average Omnidirectional Constraint 4}
\end{align}



As mentioned before, to minimize the asymptotic MD probability in (\ref{MD Probability Asymptotic}), the rank and the product of the non-zero eigenvalues of ${\bf R}$ in (\ref{Effective Channel Vector Covariance Single Path}), i.e., the rank and the product of the non-zero eigenvalues of $\tilde{\bf R}$ in (\ref{Effective Channel Vector Covariance Single Path Equivalent}), should be maximized. It is not hard to see that, for an arbitrary $\bf \Psi$, the rank of $\tilde{\bf R}$ can be maximized if and only if the $K \times K$ diagonal matrix ${\rm{diag}}\{ {\bf{a}}_1^H{{\bf{a}}_1},{\bf{a}}_2^H{{\bf{a}}_2}, \ldots ,{\bf{a}}_K^H{{\bf{a}}_K}\} $ is of full rank $K$. In addition, when $\bf \Psi$ is of full rank $K$, the product of the non-zero eigenvalues, i.e., the determinant of $\tilde{\bf R}$ follows
\begin{align}
\det (\tilde{\bf{ R}}) &= \det ({\bf{\Psi }})\det ({\rm{diag}}\{ {\bf{a}}_1^H{{\bf{a}}_1},{\bf{a}}_2^H{{\bf{a}}_2}, \ldots ,{\bf{a}}_K^H{{\bf{a}}_K}\} ) \nonumber \\
&= \det ({\bf{\Psi }})\prod\limits_{k = 1}^K {{\bf{a}}_k^H{{\bf{a}}_k}} \le \det ({\bf{\Psi }}){\Bigg( {\frac{1}{K}\sum\limits_{k = 1}^K {{\bf{a}}_k^H{{\bf{a}}_k}} } \Bigg)^K} = \det ({\bf{\Psi }}) (N_{\rm r}N_{\rm t})^K\nonumber
\end{align}
where the last equality is with (\ref{Average Omnidirectional Constraint 4}), and the equality holds if and only if all ${\bf{a}}_k^H{{\bf{a}}_k}$ have equal values, i.e.,
\begin{align}
{{\bf{v}}^H}({\theta _{\rm{t}}}){{\bf{W}}_k}{\bf{W}}_k^H{\bf{v}}({\theta _{\rm{t}}}) \cdot {{\bf{u}}^H}({\theta _{\rm{r}}}){{\bf{F}}_k}{\bf{F}}_k^H{\bf{u}}({\theta _{\rm{r}}}) = N_{\rm r}N_{\rm t}, \;\; \forall \; \theta_\mathrm{r},\theta_\mathrm{t} \in [0,1] \text{ and } \forall \; k.  \label{Omnidirectional Constraint}
\end{align}
Moreover, in (\ref{Omnidirectional Constraint}) we assume ${{\bf{v}}^H}({\theta _{\rm{t}}}){{\bf{W}}_k}{\bf{W}}_k^H{\bf{v}}({\theta _{\rm{t}}}) = {c}$ and integrate over $\theta_{\rm t}$ on the interval $[0,1]$ at both sides. The right hand side is $\int_0^1 {c\mathrm{d}{\theta _{\rm{t}}}}  = c$, and the left hand side is
\begin{align}
\int_0^1 {{{\bf{v}}^H}({\theta _{\rm{t}}}){{\bf{W}}_k}{\bf{W}}_k^H{\bf{v}}({\theta _{\rm{t}}})\mathrm{d}{\theta _{\rm{t}}}} &= \mathrm{tr}\bigg( {{{\bf{W}}_k}{\bf{W}}_k^H\int_0^1 {{\bf{v}}({\theta _{\rm{t}}}){{\bf{v}}^H}({\theta _{\rm{t}}})\mathrm{d}{\theta _{\rm{t}}}} } \bigg) = \mathrm{tr}({{\bf{W}}_k}{\bf{W}}_k^H) = {N_{\rm{t}}} \nonumber
\end{align}
where the first equality is with the fact that $\mathrm{tr}({\bf{AB}}) = \mathrm{tr}({\bf{BA}})$, the second equality is because ${\int_0^1 {{\bf{v}}({\theta _{\rm{t}}}){{{\bf{v}}^H({\theta _{\rm{t}}})}}\mathrm{d}{\theta _{\rm{t}}}} } = {\bf I}_{M_\mathrm{t}}$, which can be obtained from (\ref{Channel Model 3}) immediately, and the last equality can be obtained from (\ref{Constant Amplitude Constraint Combining Matrix}). Therefore, we have $c = N_\mathrm{t}$ and can rewrite (\ref{Omnidirectional Constraint}) as
\begin{align}
{{\bf{v}}^H}({\theta _{\rm{t}}}){{\bf{W}}_k}{\bf{W}}_k^H{\bf{v}}({\theta _{\rm{t}}}) &= {N_{\rm{t}}}, \;\; \forall \; \theta_\mathrm{t} \in [0,1] \text{ and } \forall \; k  \label{Omnidirectional Constraint Precoding Matrix} \\
{{\bf{u}}^H}({\theta _{\rm{r}}}){{\bf{F}}_k}{\bf{F}}_k^H{\bf{u}}({\theta _{\rm{r}}}) &= {N_{\rm{r}}}, \;\; \forall \; \theta_\mathrm{r} \in [0,1] \text{ and } \forall \; k. \label{Omnidirectional Constraint Combining Matrix}
\end{align}
One can observe that, conditions (\ref{Omnidirectional Constraint Precoding Matrix}) and (\ref{Omnidirectional Constraint Combining Matrix}) are more strict than (\ref{Average Omnidirectional Constraint 2}), since (\ref{Average Omnidirectional Constraint 2}) implies that omnidirectional coverage is guaranteed over $K$ synchronization time slots, while (\ref{Omnidirectional Constraint Precoding Matrix}) and (\ref{Omnidirectional Constraint Combining Matrix}) mean that omnidirectional coverage should be guaranteed at every time slot. This implies that omnidirectional transmission is superior to directional narrow beams. {\color{black}{In \cite{X.Yang2013}, the authors take cyclic delay diversity (CDD) for example. They illustrate that omnidirectional transmission is superior to directional narrow beams when considering payload data transmission. Here, we show that omnidirectional transmission is still superior to directional narrow beams when considering synchronization signals.}}

\subsubsection{The i.i.d. channel}
In this case, there are sufficiently large number of paths in (3), and we have
\begin{align}
\mathbb{E}\{ \mathrm{vec}({{\bf{H}}_k}){(\mathrm{vec}({{\bf{H}}_l}))^H}\}  = {\psi_{k,l}}{{\bf{I}}_{{M_{\rm{r}}}{M_{\rm{t}}}}} . \label{Channel Vector Covariance IID}
\end{align}
Substituting (\ref{Channel Vector Covariance IID}) into (\ref{Effective Channel Vector Covariance Subblock}) yields
\begin{align}
{{\bf{R}}_{k,l}} &= {\psi _{k,l}}({\bf{W}}_k^T \otimes {\bf{F}}_k^H)({\bf{W}}_l^* \otimes {{\bf{F}}_l}) = {\psi _{k,l}}{\bf{W}}_k^T{\bf{W}}_l^* \otimes {\bf{F}}_k^H{{\bf{F}}_l} . \label{Effective Channel Vector Covariance Subblock IID}
\end{align}
With (\ref{Effective Channel Vector Covariance Subblock IID}), we can express (\ref{Effective Channel Vector Covariance}) as
\begin{align}
{\bf{R}} &= \left[ {\begin{matrix}
{{\psi _{1,1}}{\bf{W}}_1^T{\bf{W}}_1^* \otimes {\bf{F}}_1^H{{\bf{F}}_1}}&{{\psi _{1,2}}{\bf{W}}_1^T{\bf{W}}_2^* \otimes {\bf{F}}_1^H{{\bf{F}}_2}}& \cdots &{{\psi _{1,K}}{\bf{W}}_1^T{\bf{W}}_K^* \otimes {\bf{F}}_1^H{{\bf{F}}_K}}\\
{{\psi _{2,1}}{\bf{W}}_2^T{\bf{W}}_1^* \otimes {\bf{F}}_2^H{{\bf{F}}_1}}&{{\psi _{2,2}}{\bf{W}}_2^T{\bf{W}}_2^* \otimes {\bf{F}}_2^H{{\bf{F}}_2}}& \cdots &{{\psi _{2,K}}{\bf{W}}_2^T{\bf{W}}_K^* \otimes {\bf{F}}_2^H{{\bf{F}}_K}}\\
 \vdots & \vdots & \ddots & \vdots \\
{{\psi _{K,1}}{\bf{W}}_K^T{\bf{W}}_1^* \otimes {\bf{F}}_K^H{{\bf{F}}_1}}&{{\psi _{K,2}}{\bf{W}}_K^T{\bf{W}}_2^* \otimes {\bf{F}}_K^H{{\bf{F}}_2}}& \cdots &{{\psi _{K,K}}{\bf{W}}_K^T{\bf{W}}_K^* \otimes {\bf{F}}_K^H{{\bf{F}}_K}}
\end{matrix}} \right] . \label{Effective Channel Vector Covariance IID}
\end{align}

As explained before, to minimize the asymptotic MD probability in (\ref{MD Probability Asymptotic}), the rank and the product of the non-zero eigenvalues of ${\bf R}$ in (\ref{Effective Channel Vector Covariance IID}) should be maximized. Note that the trace of the $KN_{\rm r}N_{\rm t} \times KN_{\rm r}N_{\rm t}$ matrix ${\bf R}$ in (\ref{Effective Channel Vector Covariance IID}) is
\begin{align}
\mathrm{tr}({\bf{R}}) &= \sum\limits_{k = 1}^K {\mathrm{tr}({\psi _{k,k}}{\bf{W}}_k^T{\bf{W}}_k^* \otimes {\bf{F}}_k^H{{\bf{F}}_k})} = \sum\limits_{k = 1}^K {\mathrm{tr}({\bf{W}}_k^H{{\bf{W}}_k})\mathrm{tr}({\bf{F}}_k^H{{\bf{F}}_k})} = KN_{\rm r}N_{\rm t} \nonumber
\end{align}
where the second equality is with (\ref{Channel Model 6}) and the property that $\mathrm{tr}({\bf{A}} \otimes {\bf{B}}) = \mathrm{tr}({\bf{A}})\mathrm{tr}({\bf{B}})$ and $\mathrm{tr}({\bf{A}}) = \mathrm{tr}({{\bf{A}}^T})$, and the last equality can be obtained from (\ref{Constant Amplitude Constraint Precoding Matrix}) and (\ref{Constant Amplitude Constraint Combining Matrix}) immediately. Therefore, the rank of $\bf R$ can be maximized to be $KN_{\rm r}N_{\rm t}$ and the product of non-zero eigenvalues $\bf R$, i.e., the determinant when $\bf R$ is with full rank, can be maximized to be $\det({{\bf R}}) \leq (\frac{1}{KN_{\rm r}N_{\rm t}} {\rm{tr}} ({\bf R}))^{KN_{\rm r}N_{\rm t}} = 1$ if and only if ${\bf R} = {\bf I}_{KN_{\rm r}N_{\rm t}}$. For an arbitrary channel temporal correlation matrix $\bf \Psi$ with $[{\bf \Psi}]_{k,l} = \psi_{k,l}$ in (\ref{Effective Channel Vector Covariance IID}), the equality ${\bf R} = {\bf I}_{KN_{\rm r}N_{\rm t}}$ holds  if and only if
\begin{align}
{\bf{W}}_k^T{\bf{W}}_l^* \otimes {\bf{F}}_k^H{{\bf{F}}_l} = \left\{ {\begin{matrix*}[l]
{{{\bf{I}}_{{N_{\rm{r}}}{N_{\rm{t}}}},}}&{{\text{if }} k = l}\\
{{\bf{0}},}&{\text{otherwise}.}
\end{matrix*}} \right. \nonumber
\end{align}
This implies that at each synchronization time slot, both of the precoding and combining matrices should be unitary, i.e.,
\begin{align}
{\bf{W}}_k^H{{\bf{W}}_k} &= {{\bf{I}}_{{N_{\rm{t}}}}}, \;\; \forall \; k  \label{Orthogonal Constraint Precoding Matrix} \\
{\bf{F}}_k^H{{\bf{F}}_k} &= {{\bf{I}}_{{N_{\rm{r}}}}}, \;\; \forall \; k. \label{Orthogonal Constraint Combining Matrix}
\end{align}
Moreover, at two different synchronization time slots, either the precoding matrices or the combining matrices at these two time slots should be orthogonal to each other, i.e.,
\begin{align}
{{\bf{W}}_k^H{{\bf{W}}_l} = {\bf{0}} \;\; \text{or} \;\; {\bf{F}}_k^H{{\bf{F}}_l} = {\bf{0}},} \;\; \forall \; k \neq l. \label{Orthogonal Constraint Precoding Combining Matrix}
\end{align}

\subsection{FA Probability}

In this subsection, we mainly investigate the effect of the recoding and combining matrices to the FA probability. First, we present the following theorem.

\emph{Theorem 4:} The FA probability (\ref{FA Probability Definition}) can be expressed as
\begin{align}
P_{\mathrm{FA}} = \mathbb{P}\Bigg\{ {\frac{{\sum\nolimits_{k = 1}^K {\| { {{\bf{Z}}_{k,2}}} \|_{\rm{F}}^2} }}{{\sum\nolimits_{k = 1}^K {\| {{{\bf{Z}}_{k,1}}} \|_{\rm{F}}^2} }} > \frac{\gamma}{1-\gamma} } \Bigg\} \label{FA Probability}
\end{align}
where ${\bf Z}_{k,1} \in \mathbb{C}^{N_\mathrm{r} \times (L-N_\mathrm{t})}$ and ${\bf Z}_{k,2} \in \mathbb{C}^{N_\mathrm{r} \times N_\mathrm{t}}$ are both with i.i.d. $\mathcal{CN}(0,\nu)$ entries and they are independent of each other.
\begin{IEEEproof}
See Appendix B.
\end{IEEEproof}

It can be observed that neither the precoding matrix nor the combining matrix affects the FA probability (\ref{FA Probability}). Moreover, note that the left hand side in $\mathbb{P}(\cdot)$ in (\ref{FA Probability}) follows the standard F-distribution. Therefore, according to the cumulative distribution function (CDF) of F-distribution, a closed-form expression of (\ref{FA Probability}) can be expressed as \cite{C.Walck2007}
\begin{align}
{P_{{\rm{FA}}}} = {( {1 - \gamma } )^{KL{N_{\rm{r}}} - 1}}\sum\limits_{m = 0}^{K{N_{\rm{r}}}{N_{\rm{t}}} - 1} { \binom{KL{N_{\rm{r}}} - 1}{m} {{\bigg( {\frac{\gamma }{{1 - \gamma }}} \bigg)^m}}} . \label{FA Probability Closed Form}
\end{align}

As a conclusion of this section, we have analyzed the effect of the precoding and combining matrices $\{{\bf W}_k,{\bf F}_k\}_{k=1}^K$ to the synchronization performance. It is shown that $\{{\bf W}_k,{\bf F}_k\}_{k=1}^K$ should satisfy (\ref{Omnidirectional Constraint Precoding Matrix}) and (\ref{Omnidirectional Constraint Combining Matrix}) to  minimize the asymptotic MD probability under the single-path channel, and satisfy (\ref{Orthogonal Constraint Precoding Matrix}), (\ref{Orthogonal Constraint Combining Matrix}), and (\ref{Orthogonal Constraint Precoding Combining Matrix}) to minimize the asymptotic MD probability under the i.i.d. channel. Moreover, it is shown that $\{{\bf W}_k,{\bf F}_k\}_{k=1}^K$ do not affect the FA probability.

\section{Design of Precoding and Combining Matrices}

In Section III, it is shown that the precoding and combining matrices $\{{\bf W}_k,{\bf F}_k\}_{k=1}^{K}$ should satisfy conditions (\ref{Constant Amplitude Constraint Precoding Matrix}), (\ref{Constant Amplitude Constraint Combining Matrix}), and (\ref{Average Omnidirectional Constraint 2}) to guarantee the requirements of constant amplitude for all the entries therein and omnidirectional coverage over total $K$ synchronization time slots. In Section IV, it is shown that $\{{\bf W}_k,{\bf F}_k\}_{k=1}^{K}$ should satisfy conditions (\ref{Omnidirectional Constraint Precoding Matrix}), (\ref{Omnidirectional Constraint Combining Matrix}), (\ref{Orthogonal Constraint Precoding Matrix}), (\ref{Orthogonal Constraint Combining Matrix}), and (\ref{Orthogonal Constraint Precoding Combining Matrix}) to minimize the asymptotic MD probability under the single-path channel and the i.i.d. channel. Note that (\ref{Average Omnidirectional Constraint 2}) can be satisfied automatically as long as (\ref{Omnidirectional Constraint Precoding Matrix}) and (\ref{Omnidirectional Constraint Combining Matrix}) have been satisfied, hence it can be ignored. In this section, we will investigate how to design $\{{\bf W}_k,{\bf F}_k\}_{k=1}^{K}$ to satisfy conditions (\ref{Constant Amplitude Constraint Precoding Matrix}), (\ref{Constant Amplitude Constraint Combining Matrix}), (\ref{Omnidirectional Constraint Precoding Matrix}), (\ref{Omnidirectional Constraint Combining Matrix}), (\ref{Orthogonal Constraint Precoding Matrix}), (\ref{Orthogonal Constraint Combining Matrix}), and (\ref{Orthogonal Constraint Precoding Combining Matrix}) simultaneously.

First, to simplify the problem, we consider the case with $K = 1$, then condition (\ref{Orthogonal Constraint Precoding Combining Matrix}) can be ignored temporarily. Since conditions (\ref{Constant Amplitude Constraint Precoding Matrix}), (\ref{Omnidirectional Constraint Precoding Matrix}), and (\ref{Orthogonal Constraint Precoding Matrix}) are similar to (\ref{Constant Amplitude Constraint Combining Matrix}), (\ref{Omnidirectional Constraint Combining Matrix}), and (\ref{Orthogonal Constraint Combining Matrix}), respectively, the design of the precoding matrix and the design of the combining matrix will also be similar. Therefore in the following contents, we mainly take precoding for example to describe the design procedure. We rewrite conditions (\ref{Constant Amplitude Constraint Precoding Matrix}), (\ref{Omnidirectional Constraint Precoding Matrix}), and (\ref{Orthogonal Constraint Precoding Matrix}) as
\begin{align}
{[{\bf{W}}]_{m,n}} &= \frac{1}{{\sqrt M }}, \;\; \forall \; m,n \label{Constant Amplitude Constraint Matrix} \\
{{\bf{v}}^H}(\theta ){\bf{W}}{{\bf{W}}^H}{\bf{v}}(\theta ) &= 1, \;\; \forall \; \theta  \in [0,1] \label{Omnidirectional Constraint Matrix} \\
{{\bf{W}}^H}{\bf{W}} &= {{\bf{I}}_N} \label{Orthogonal Constraint Matrix}
\end{align}
where we have omitted the subscripts $k$ and $\rm t$ for notational simplicity. We need to design matrix ${\bf W}\in \mathbb{C}^{M\times N}$ to satisfy (\ref{Constant Amplitude Constraint Matrix}), (\ref{Omnidirectional Constraint Matrix}), and (\ref{Orthogonal Constraint Matrix}) simultaneously.

Letting ${\bf{W}} = [{{\bf{w}}_1},{{\bf{w}}_2}, \ldots ,{{\bf{w}}_N}]$ and ${{\bf{w}}_n} = {[{w_{1,n}},{w_{2,n}}, \ldots ,{w_{M,n}}]^T}$, we can rewrite (\ref{Omnidirectional Constraint Matrix}) as
\begin{align}
{{\bf{v}}^H}(\theta ){\bf{W}}{{\bf{W}}^H}{\bf{v}}(\theta ) &= \sum\limits_{n = 1}^N {{{\left| {\sum\limits_{m = 1}^M {{w_{m,n}}{e^{j2\pi (m - 1)\theta }}} } \right|}^2}} = \sum\limits_{l =  - M + 1}^{M - 1} { {\sum\limits_{n = 1}^N {{r_{l,n}}} } {e^{j2\pi l\theta }}} \label{Relation Between Matrix and Sequence}
\end{align}
where
\begin{align}
{r_{l,n}} = \left\{ {\begin{matrix*}[l]
{\sum\limits_{m = 1}^M {{w_m}w_{m + l}^*} ,}&{l = 0,1, \ldots ,M - 1}\\
{r_{ - l,n}^*,}&{l =  - M + 1, \ldots ,- 1}
\end{matrix*}} \right. \nonumber
\end{align}
represents the aperiodic autocorrelation function of ${\bf w}_n$, if we regard vector ${\bf w}_n$ as a sequence of length $M$. Substituting (\ref{Relation Between Matrix and Sequence}) into (\ref{Omnidirectional Constraint Matrix}) yields
\begin{align}
\sum\limits_{l =  - M + 1}^{M - 1} {\sum\limits_{n = 1}^N {{r_{l,n}}} {e^{j2\pi l\theta }}}  = 1, \;\; \forall \; \theta  \in [0,1]. \label{Omnidirectional Constraint Sequence}
\end{align}
According to the property of Fourier transform, we know that (\ref{Omnidirectional Constraint Sequence}) holds if and only if
\begin{align}
{\sum\limits_{n = 1}^N {{r_{l,n}}} } = \delta_l, \label{Sequence Delta Function}
\end{align}
i.e., the $N$ respective aperiodic autocorrelation functions of $\{{\bf w}_n\}_{n=1}^N$ sum to a Kronecker delta function. We need to design such $N$ vectors $\{{\bf w}_n\}_{n=1}^N$ satisfying (\ref{Sequence Delta Function}), (\ref{Constant Amplitude Constraint Matrix}), and (\ref{Orthogonal Constraint Matrix}) simultaneously.

When $N=1$, i.e., there is only one vector ${\bf w}$ and we have omitted the subscript $n$ for notational simplicity, condition (\ref{Sequence Delta Function}) implies that the aperiodic autocorrelation function of ${\bf w}$ is a Kronecker delta function. However, it is known that the aperiodic autocorrelation function of ${\bf w}$ is a Kronecker delta function if and only if there is only one non-zero entry in $\bf w$. This implies that condition (\ref{Constant Amplitude Constraint Matrix}) cannot be satisfied simultaneously since it requires constant amplitude for all the $M$ entries in ${\bf w}$. Therefore, when employing single-stream precoding at the BS side or single-stream combining at the UT side, it is impossible to generate perfect omnidirectional beam while guaranteeing constant amplitude for all the entries in the precoding or combining matrix. In our previous studies \cite{X.Meng2016,X.Meng2017}, we proposed to use ZC sequences as the precoding vector. Note that this approach can only guarantee equal transmission power in discrete angles, i.e., quasi-omnidirectional.

However, when $N=2$, the situation begins to change. We will show that in this case, conditions (\ref{Constant Amplitude Constraint Matrix}), (\ref{Omnidirectional Constraint Matrix}), and (\ref{Orthogonal Constraint Matrix}) can be satisfied simultaneously. Before proceeding, we introduce the Golay complementary pair, which was first proposed by M. Golay in the context of infrared spectrometry \cite{M.J.E.Golay1949}. A Golay complementary pair consists of two binary sequences, i.e., vectors, with the same length and satisfying the complementary property that the respective aperiodic autocorrelation functions of these two vectors sum to a Kronecker delta function \cite{M.J.E.Golay1961}. There are many approaches to construct Golay complementary pairs, including direct constructions and recursive constructions \cite{M.J.E.Golay1961,M.J.E.Golay1977,J.A.Davis1999}. Here, we introduce the Golay-Rudin-Shapiro recursion construction \cite{M.J.E.Golay1961}, where the two vectors ${\bf p}_{M,1}$ and ${\bf p}_{M,2}$ of length $M$ are recursively constructed with the two vectors ${\bf p}_{M/2,1}$ and ${\bf p}_{M/2,2}$ of length $M/2$ as
\begin{align}
{{\bf{p}}_{M,1}} &= \left[ {\begin{matrix}
{{{\bf{p}}_{M/2,1}}}\\
{{{\bf{p}}_{M/2,2}}}
\end{matrix}} \right], \;\; {{\bf{p}}_{M,2}} = \left[ {\begin{matrix}
{{{\bf{p}}_{M/2,1}}}\\
{ - {{\bf{p}}_{M/2,2}}}
\end{matrix}} \right] \label{Golay Complementary Pair}
\end{align}
with initial vectors ${{\bf{p}}_{1,1}} = {{\bf{p}}_{1,2}} = [1]$. Moreover, it is shown that the respective aperiodic autocorrelation functions of ${\bf p}_{M,1}$ and ${\bf p}_{M,2}$ constructed with (\ref{Golay Complementary Pair}) sum to a Kronecker delta function \cite{M.J.E.Golay1961}.


With the Golay complementary pair, the design is straightforward as long as we let ${\bf W} = \frac{1}{\sqrt{M}} [{\bf p}_{M,1},{\bf p}_{M,2}]$. Since both ${\bf p}_{M,1}$ and ${\bf p}_{M,2}$ are binary vectors, condition (\ref{Constant Amplitude Constraint Matrix}) can be satisfied. The complementary property of ${\bf p}_{M,1}$ and ${\bf p}_{M,2}$ lets condition (\ref{Omnidirectional Constraint Sequence}), i.e., (\ref{Omnidirectional Constraint Matrix}) be satisfied. Moreover, ${\bf p}_{M,1}$ and ${\bf p}_{M,2}$ are also orthogonal to each other, hence condition (\ref{Orthogonal Constraint Matrix}) can be satisfied.

Then consider the case with $N > 2$ and $N$ is even, we can use the Golay-Hadamard matrix \cite{X.Huang2002}, a generalization of the Golay complementary pair, for the design. An $M \times M$ Golay-Hadamard matrix ${\bf P}_M$ can be recursively constructed as
\begin{align}
{{\bf{P}}_M} &= \left[ {\begin{matrix}
{{{\bf{P}}_{M/2}}}&{{{\bf{P}}_{M/2}}}\\
{{{\tilde{\bf{P}}}_{M/2}}}&{ - {{\tilde{\bf{P}}}_{M/2}}}
\end{matrix}} \right], \;\; {{\tilde{\bf{P}}}_M} = \left[ {\begin{matrix}
{{{\bf{P}}_{M/2}}}&{{{\bf{P}}_{M/2}}}\\
{ - {{\tilde{\bf{P}}}_{M/2}}}&{{{\tilde{\bf{P}}}_{M/2}}}
\end{matrix}} \right] \label{Golay Hadamard Matrix}
\end{align}
with initial matrices ${{\bf{P}}_1} = {{\tilde{\bf{P}}}_1} = [1]$. For a Golay-Hadamard matrix ${\bf P}_M$ constructed from (\ref{Golay Hadamard Matrix}), it can be proved that the $n$th column and the $(M/2+n)$th column of ${\bf P}_M$ constitute a Golay complementary pair for $n = 1,2,\ldots,M/2$ \cite{X.Huang2002}. By using this complementary property, the design of matrix ${\bf W}$ can be completed  by selecting the $n_l$th column and the corresponding $(M/2+n_l)$th column of ${\bf P}_M$ for $l = 1,2,\ldots,N/2$, totally $N$ columns, as the columns of $\bf W$. In mathematical expression we have
\begin{align}
{\bf{W}} = \frac{1}{\sqrt{M}} \big[{[{{\bf{P}}_M}]_{:,{n_1}}},{[{{\bf{P}}_M}]_{:,{n_1} + M/2}}, \ldots ,{[{{\bf{P}}_M}]_{:,{n_{N/2}}}},{[{{\bf{P}}_M}]_{:,{n_{N/2}} + M/2}}\big] \nonumber
\end{align}
where ${n_1},{n_2}, \ldots ,{n_{N/2}} \in \{ 1,2, \ldots ,M/2\} $ need to be different from each other. Obviously, with the above constructed $\bf W$, conditions (\ref{Constant Amplitude Constraint Matrix}) and (\ref{Omnidirectional Constraint Matrix}) can be satisfied. In addition, since ${\bf P}_M$ itself is a unitary matrix, all the $N$ columns of $\bf W$ will be orthogonal to each other. Hence condition (\ref{Orthogonal Constraint Matrix}) can also be satisfied.



Finally, we consider the general case with $K \geq 1$ and present the design of the precoding and combining matrices $\{ {\bf W}_k,{\bf F}_k \}_{k=1}^K$. Note that each ${\bf W}_k$ is of size $M_{\mathrm{t}} \times N_{\mathrm{t}}$ and each ${\bf F}_k$ is of size $M_{\mathrm{r}} \times N_{\mathrm{r}}$. Let both $M_\mathrm{t}$ and $M_\mathrm{r}$ be an integer power of $2$, and both $N_\mathrm{t}$ and $N_\mathrm{r}$ be an integer multiple of $2$. Then ${\bf W}_k$ and ${\bf F}_k$ can be constructed as
\begin{align}
{{\bf{W}}_k} &= \frac{1}{{\sqrt {{M_{\rm{t}}}} }}\big[{[{{\bf{P}}_{{M_{\rm{t}}}}}]_{:,{n_{{\rm{t}},k,1}}}},{[{{\bf{P}}_{{M_{\rm{t}}}}}]_{:,{n_{{\rm{t}},k,1}} + {M_{\rm{t}}}/2}}, \ldots ,{[{{\bf{P}}_{{M_{\rm{t}}}}}]_{:,{n_{{\rm{t}},k,{N_{\rm{t}}}/2}}}},{[{{\bf{P}}_{{M_{\rm{t}}}}}]_{:,{n_{{\rm{t}},k,{N_{\rm{t}}}/2}} + {M_{\rm{t}}}/2}}\big] \label{Precoding Matrix Design} \\
{{\bf{F}}_k} &= \frac{1}{{\sqrt {{M_{\rm{r}}}} }}\big[{[{{\bf{P}}_{{M_{\rm{r}}}}}]_{:,{n_{{\rm{r}},k,1}}}},{[{{\bf{P}}_{{M_{\rm{r}}}}}]_{:,{n_{{\rm{r}},k,1}} + {M_{\rm{r}}}/2}}, \ldots ,{[{{\bf{P}}_{{M_{\rm{r}}}}}]_{:,{n_{{\rm{r}},k,{N_{\rm{r}}}/2}}}},{[{{\bf{P}}_{{M_{\rm{r}}}}}]_{:,{n_{{\rm{r}},k,{N_{\rm{r}}}/2}} + {M_{\rm{r}}}/2}}\big] \label{Combining Matrix Design}
\end{align}
where both ${\bf P}_{M_\mathrm{t}}$ and ${\bf P}_{M_\mathrm{r}}$ are Golay-Hadamard matrices constructed according to (\ref{Golay Hadamard Matrix}), for each $k$, ${n_{{\rm{t}},k,1}},{n_{{\rm{t}},k,2}},\; \ldots ,{n_{{\rm{t}},k,{N_{\rm{t}}}/2}} \in \{ 1,2,\ldots,M_\mathrm{t}/2 \}$ need to be different from each other, and  ${n_{{\rm{r}},k,1}},{n_{{\rm{r}},k,2}}, \ldots ,{n_{{\rm{r}},k,{N_{\rm{r}}}/2}} \in \{ 1,2,\ldots,M_\mathrm{r}/2 \}$ need to be different from each other. In addition, to satisfy condition (\ref{Orthogonal Constraint Precoding Combining Matrix}), for each $k \neq l$, it should be satisfied that $\{{n_{{\rm{t}},k,1}},{n_{{\rm{t}},k,2}}, \ldots ,{n_{{\rm{t}},k,{N_{\rm{t}}}/2}} \} \cap \{{n_{{\rm{t}},l,1}},{n_{{\rm{t}},l,2}}, \ldots ,{n_{{\rm{t}},l,{N_{\rm{t}}}/2}} \} = \varnothing$ or $\{{n_{{\rm{r}},k,1}},{n_{{\rm{r}},k,2}}, \ldots ,{n_{{\rm{r}},k,{N_{\rm{t}}}/2}} \} \cap \{{n_{{\rm{r}},l,1}},{n_{{\rm{r}},l,2}}, \ldots ,{n_{{\rm{r}},l,{N_{\rm{t}}}/2}} \} = \varnothing$.

\section{Numerical Results}

In this section, we present numerical simulations to evaluate the performance of mmWave massive MIMO synchronization with the proposed omnidirectional precoding and combining approach. The BS has $M_\mathrm{t} = 64$ antennas and the UT has $M_\mathrm{r} = 16$ antennas. The number of channel paths in (\ref{Channel Model 1}) is set as $P = 1$ or $P = 4$. For both of these two cases, the arrival and departure angles $\theta_{\mathrm{r},p}$ and $\theta_{\mathrm{t},p}$ of each path in (\ref{Channel Model 1}) randomly take values in $[0,1]$, and the average gain of each path in (\ref{Channel Model 4}) is $\beta_p = 1/P$. The temporal correlation coefficient in (\ref{Channel Model 5}) is generated as $\psi_{k,l} = J_0(2\pi f_\mathrm{d} T_\mathrm{s} |k-l|)$ with $f_\mathrm{d} = v f_\mathrm{c} / c$ \cite{J.Choi2014,J.He2014,G.C.Alexandropoulos2016}, where $J_0(\cdot)$ denotes the Bessel function of the first kind, $v = 30$ km/h denotes the velocity of the UT, $f_\mathrm{c} = 30$ GHz denotes the carrier frequency, $c = 3\times 10^{8}$ m/s denotes the speed of light, and {\color{black}{$T_\mathrm{s} = 0.5$ ms denotes the time interval between two adjacent synchronization time slots.}} The length of the synchronization signal ${\bf X}_k$ is $L = 64$. {\color{black}{We simulate total $500$ drops to generate the arrival and departure angles $\theta_{\mathrm{r},p}$ and $\theta_{\mathrm{t},p}$ of each path in $[0,1]$ randomly. In each drop, the arrival and departure angles are fixed, and only fast fading are considered. The total number of time slots is $10000$ for each drop. The final performance curves are the average results of drops and time slots.}}

First, we consider the case with $K = 1$, i.e., the UT utilizes the received signal at $K = 1$ time slot to synchronize with the BS. This corresponds to the scenario with a short latency time and a relatively low success probability for initial synchronization. We compare the performance between three different precoding and combining approaches, including: 1) omnidirectional precoding and omnidirectional combining proposed in this paper; 2) quasi-omnidirectional precoding and omnidirectional combining; 3) random precoding and random combining. For Approach 1, we let $N_\mathrm{t}^{(1)} = N_\mathrm{r}^{(1)} = 2$. The precoding and combining matrices are generated according to (\ref{Precoding Matrix Design}) and (\ref{Combining Matrix Design}), where $n_{\mathrm{t},k,1} = n_{\mathrm{r},k,1} = 1$, $n_{\mathrm{t},k,2} = M_\mathrm{t}/2$, $n_{\mathrm{r},k,2} = M_\mathrm{r}/2$, and $k = 1 $ since $K=1$. {\color{black}{For Approach 2, we let $N_\mathrm{t}^{(2)} = 1$ and $N_\mathrm{r}^{(2)} = 2$. The precoding vector is set as a ZC sequence of length $64$, and the combining matrix is the same as that in Approach 1. For Approach 3, we let $N_\mathrm{t}^{(3)} = N_\mathrm{r}^{(3)} = 1$. All the entries in the precoding and combining vectors have constant amplitudes and i.i.d. $\mathcal{U}(0,2\pi)$ phases.}} To guarantee fair comparison, we let the FA probabilities of all these three approaches be equal to $10^{-4}$. This can be achieved by letting $P_{\mathrm{FA}} = 10^{-4}$ in (\ref{FA Probability Closed Form}) and then obtaining the corresponding threshold values $\gamma^{(1)},\gamma^{(2)},\gamma^{(3)}$ for these three approaches respectively.

The MD probabilities with respect to the SNR value for the above three approaches obtained from (\ref{MD Probability Simple}) are presented in Figs. 2 and 3. {\color{black}{It can be observed that Approach 1, denoted as ``omni precoding, omni combining'', has the best performance. This is because it can guarantee perfect omnidirectional coverage at
both of the BS and UT sides, hence there is no transmission power fluctuation with respect to spatial angle directions. Since a ZC sequence is used as the precoding vector and random sequences are used as the precoding and combining vectors therein, Approaches 2 and 3, denoted as ``omni combining, quasi-omni precoding'' and ``random combining, random
precoding'', have transmission power nulls and fluctuation in spatial angle directions. This will lead to performance loss when the nulls or the angle directions with relatively low power align with the channel paths.}} Moreover, for the case with $P=4$ in Fig. 3, the performance curve of Approach 1 has a larger slope than the other two approaches. This is because the maximum achievable diversity order of Approach 1 is $N_\mathrm{r}^{(1)} N_\mathrm{t}^{(1)} K = 4$. When the actual channel has $P=4$ paths, this diversity order can be exploited. Also note that the maximum achievable diversity orders of the other two approaches are $N_\mathrm{r}^{(2)} N_\mathrm{t}^{(2)} K = 2$ and $N_\mathrm{r}^{(3)} N_\mathrm{t}^{(3)} K = 1$, respectively. {\color{black}{The relatively high SNR in Figs. 2 and 3 is because the UT needs to synchronize with the BS in a very short latency time ($KT = 0.5$ ms). This will obviously lead to poor synchronization performance. These two figures are mainly used to demonstrate that in the scenario with a short latency time and a relatively low success probability for initial synchronization, our proposed approach is superior to other existing approaches. In Figs. 4 and 5, where the latency time is relatively large ($KT = 32$ ms), the resulting SNR will be relatively low.}}

\begin{figure}[htbp]
\centering
\includegraphics[width=0.5\columnwidth]{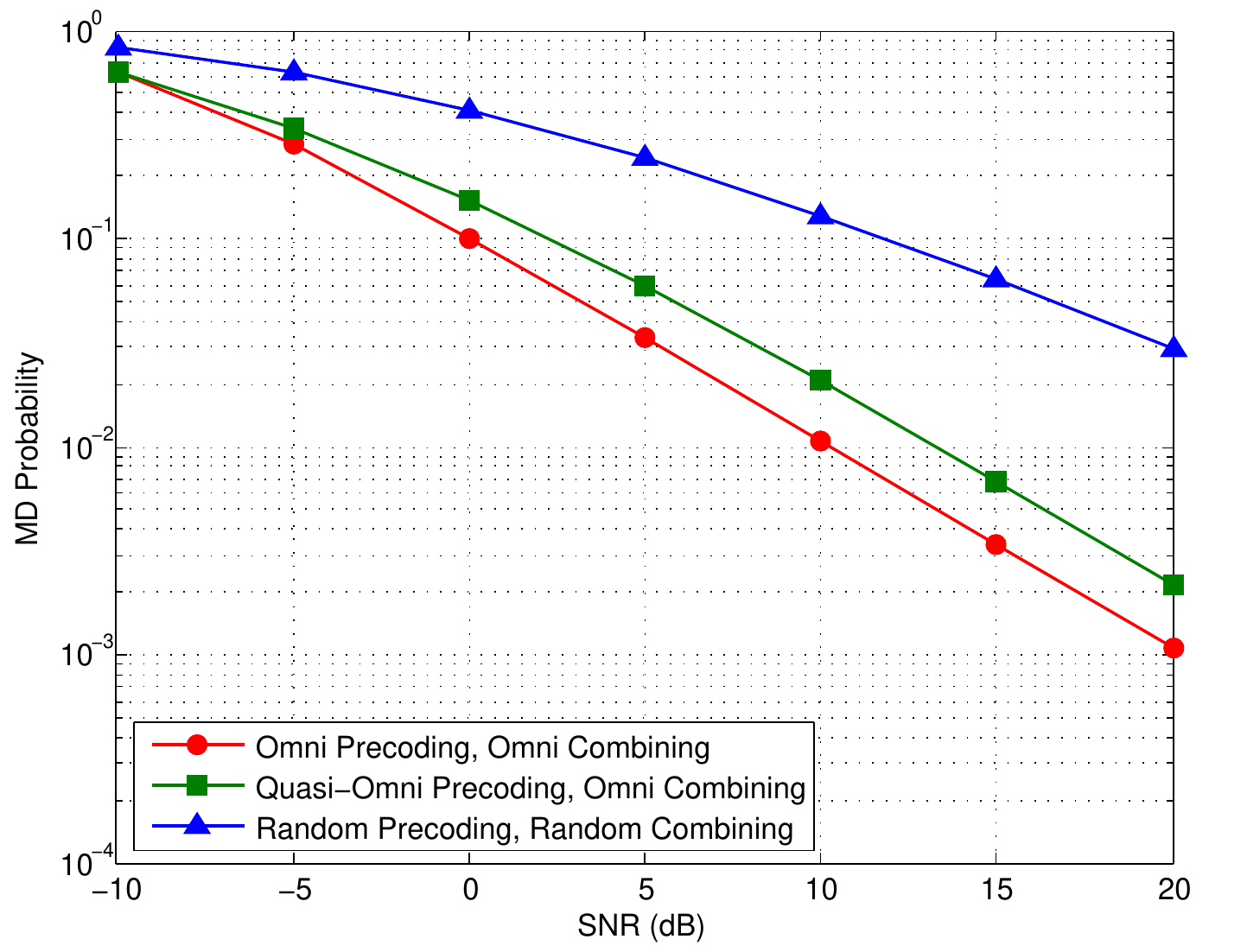}
\caption{Comparison of MD probability between different precoding and combining approaches, where $K = 1$ and $P = 1$.}
\end{figure}

\begin{figure}[htbp]
\centering
\includegraphics[width=0.5\columnwidth]{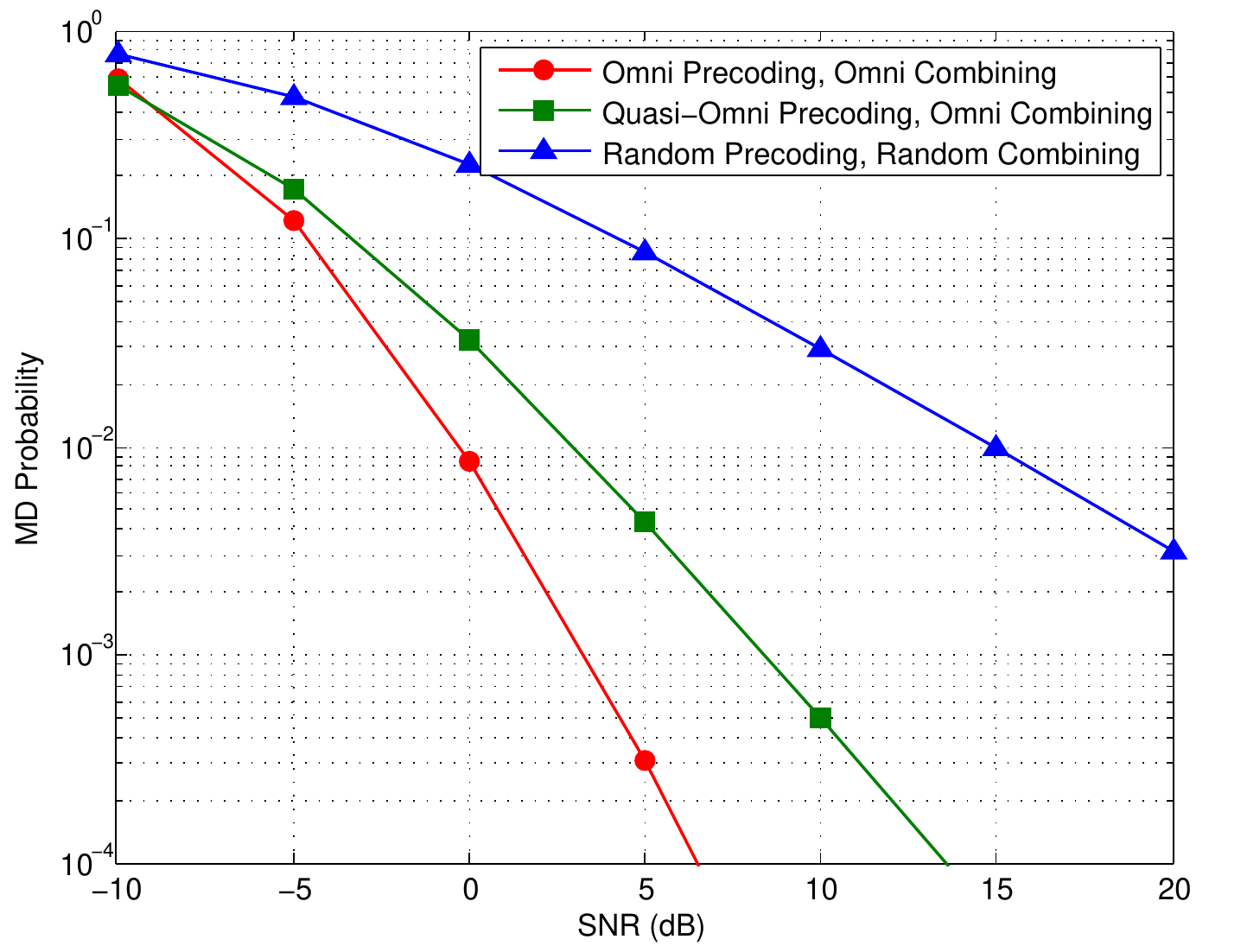}
\caption{Comparison of MD probability between different precoding and combining approaches, where $K = 1$ and $P = 4$.}
\end{figure}

\begin{figure}[htbp]
\centering
\includegraphics[width=0.5\columnwidth]{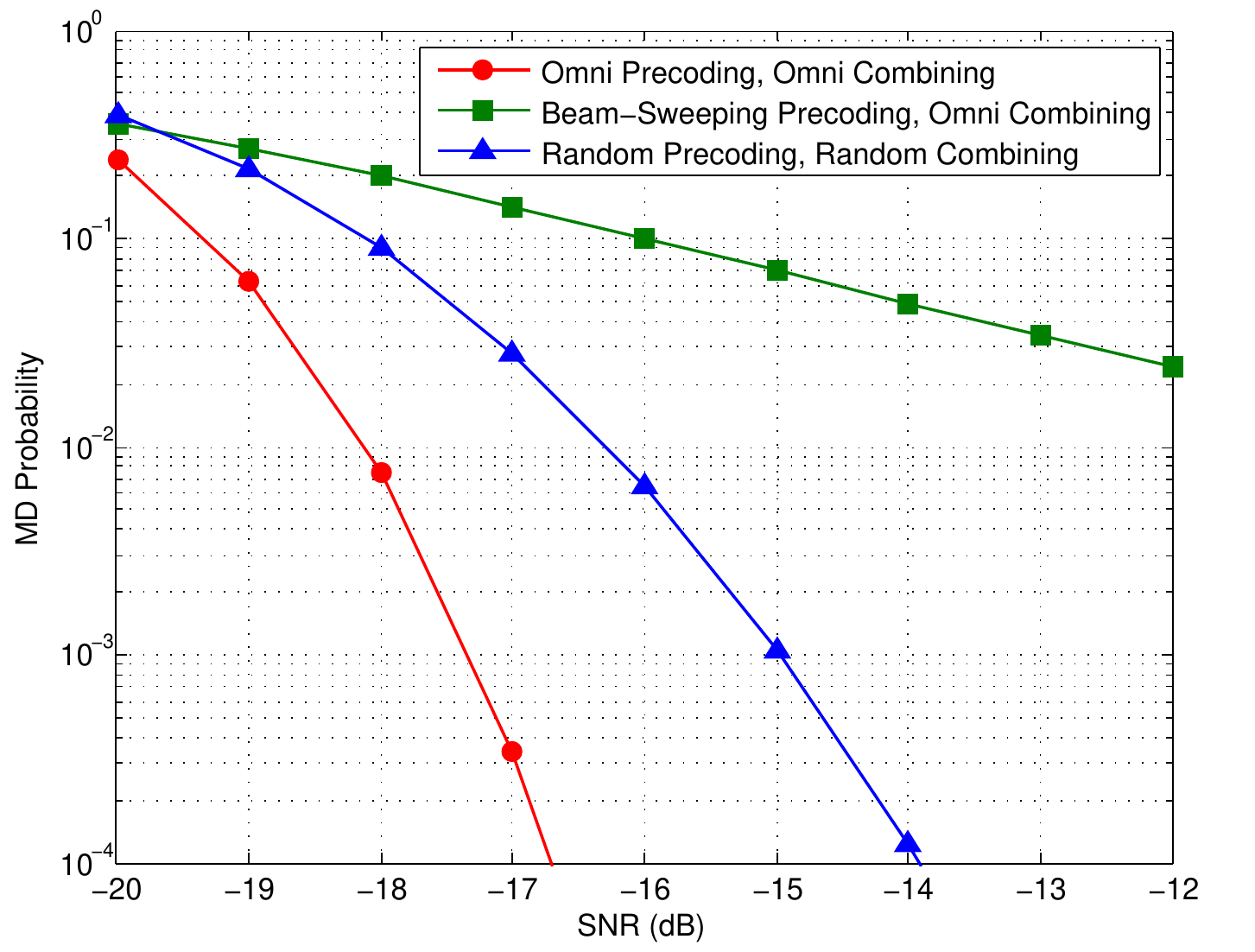}
\caption{Comparison of MD probability between different precoding and combining approaches, where $K = 64$ and $P = 1$.}
\end{figure}

\begin{figure}[htbp]
\centering
\includegraphics[width=0.5\columnwidth]{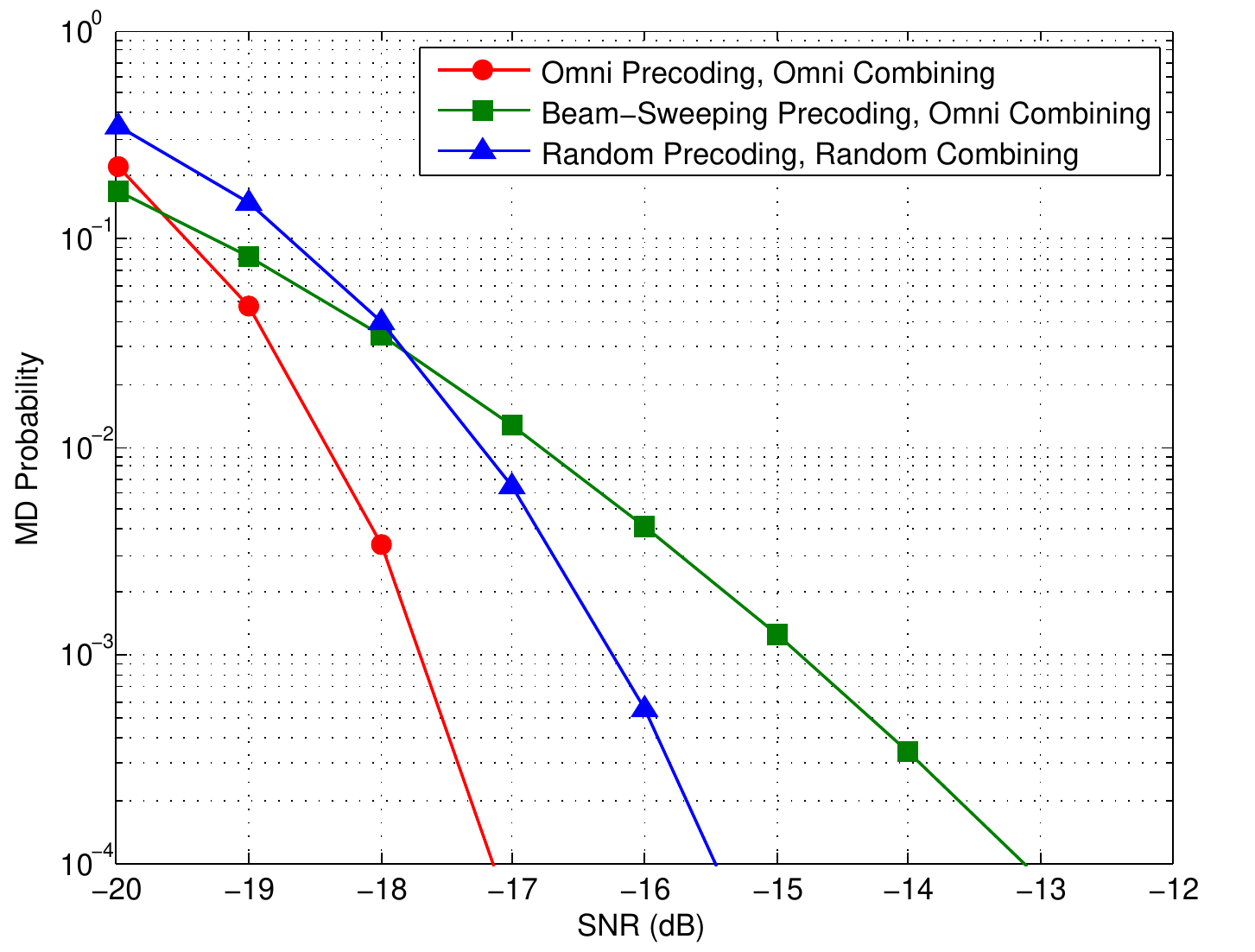}
\caption{Comparison of MD probability between different precoding and combining approaches, where $K = 64$ and $P = 4$.}
\end{figure}

Then, we consider the case with $K = 64$, i.e., the UT utilizes the received signals at $K = 64$ time slots to synchronize with the BS. This corresponds to the scenario with a long latency time and a relatively high success probability of initial synchronization. We compare the performance between three different precoding and combining approaches, including: 1) omnidirectional precoding and omnidirectional combining proposed in this paper; 2) beam-sweeping precoding and omnidirectional combining; 3) random precoding and random combining. For Approach 1, we let $N_\mathrm{t}^{(1)} = N_\mathrm{r}^{(1)} = 2$, and the precoding and combining matrices are generated according to (\ref{Precoding Matrix Design}) and (\ref{Combining Matrix Design}), where $n_{\mathrm{t},k,1} = ((k))_{M_\mathrm{t}/2}$, $n_{\mathrm{t},k,2} = ((k))_{M_\mathrm{t}/2}+M_\mathrm{t}/2$, $n_{\mathrm{r},k,1} = ((k))_{M_\mathrm{r}/2}$, $n_{\mathrm{r},k,2} = ((k))_{M_\mathrm{r}/2}+M_\mathrm{r}/2$, and $k = 1,2,\ldots, 64$ since $K=64$. For Approach 2, we let $N_\mathrm{t}^{(2)} = 1$ and $N_\mathrm{r}^{(2)} = 2$. The combining matrix is the same with that in Approach 1, and the precoding vector is set as the columns of a $64 \times 64$ discrete Fourier transform (DFT) matrix, i.e., $\frac{1}{\sqrt{64}}[1,e^{j2\pi k/64},\ldots,e^{j2\pi 63k/64}]^T$ for $k = 1,2,\ldots, 64$. For Approach 3, we let $N_\mathrm{t}^{(3)} = N_\mathrm{r}^{(3)} = 1$, and all the entries in the precoding and combining vectors at each time slot have constant amplitudes and i.i.d. $\mathcal{U}(0,2\pi)$ phases. The FA probabilities of all these three approaches are set as $10^{-4}$.

The MD probabilities with respect to the SNR value for the above three approaches obtained from (\ref{MD Probability Simple}) are presented in Figs. 4 and 5. In Fig. 4, where the number of channel paths is $P=1$, it can be observed that Approach 2, denoted as ``beam-sweeping precoding, omni combining'', has a very poor performance. This is because it use narrow beams towards different spatial angle directions in total $K=64$ time slots to guarantee omnidirectional coverage at the BS side. When there is only $1$ channel path, the narrow beam could align with this path only at one time slot. At the other $K-1=63$ time slots, the UT will receive little signal power. This implies that the effective channel between the BS and UT over the total $K=64$ time slots will include $1$ very strong component and $63$ nearly zero components. Hence its time diversity order is only $1$. As a comparison, Approach 1, denoted as ``omni precoding, omni combining'', can guarantee omnidirectional coverage at every time slot. Therefore, the effective channel over the total $K=64$ time slots will include $64$ weak components, and time diversity order $64$ can be exploited. In addition, for Approach 3, denoted as ``random precoding, random combining'', the random precoding and combining vectors therein generate neither a perfect omnidirectional beam as Approach 1, nor a single narrow beam as Approach 2. Its performance is between those of Approaches 1 and 2. In Fig. 5, where the number of channel paths is $P=4$, it can be observed that Approach 2 shows a better performance (larger slope) than that in Fig. 4. This is because when there are $P=4$ paths having different spatial angles, the narrow beam could probably align with one of these $4$ paths at $4$ time slots. Therefore, the effective channel between the BS and UT over the total $K=64$ time slots will include $4$ relatively strong components and $60$ nearly zero components. Hence time diversity order $4$ can be obtained. Moreover, it has to be noted that in Section IV-B when we analyze the effect of the precoding and combining matrices to the MD probability, we use the asymptotic MD probability (\ref{MD Probability Asymptotic}) obtained at relatively high SNR (low MD probability) regime. This means that our proposed Approach 1 is preferable when the MD probability is low. In Fig. 5, we notice that when the SNR value is $-20$ dB, i.e., the MD probability is high (greater than $10^{-1}$), the performance of Approach 2 is better than that of Approach 1. However, in practice, the MD probability should be low enough. Otherwise the synchronization may fail and the communication system may not work. Therefore, the low MD probability regime is relevant to practical applications, and our proposed Approach 1 shows significant performance gain in this regime.

\section{Conclusions}

We have proposed an omnidirectional precoding and omnidirectional combining approach for mmWave massive MIMO synchronization. We demonstrated two basic requirements for the precoding and combining matrices, including that all the entries therein should have constant amplitude, and the transmission power averaged over the total $K$ time slots should be constant for any spatial direction. Then, by utilizing the GLRT based synchronization detector, we analyzed the effect of the precoding and combining matrices to the MD probability and the FA probability, respectively, and present the corresponding conditions that should be satisfied. It is shown that, both of the precoding and combining matrices should guarantee perfectly omnidirectional coverage at each time slot, to minimize the asymptotic MD probability under the single-path channel. Since such omnidirectional precoding matrices and omnidirectional combining matrices exist only when both of the number of transmit streams and the number of receive streams are equal to or greater than two, we utilized Golay complementary pairs and Golay-Hadamard matrices to design the precoding and combining matrices. Simulation results verify the effectiveness of the propose approach.

\appendices

\section{Proof of Theorem 1}

With (\ref{Sync Signal Model}) and under hypothesis $\mathcal{H}_1$, the logarithm of the probability density function (PDF) of the observed signal over $K$ synchronization time slots can be expressed as
\begin{align}
\ln f( {{\bf{Y}}(\tau)|{\mathcal{H}_1},{\bf{G}},{\nu}} ) &=  - KL{N_{\rm{r}}}\ln ( {\pi {\nu}} ) - L\sum\limits_{k = 1}^K {\ln \det ( {{\bf{F}}_k^H{{\bf{F}}_k}} )} \nonumber \\
& - \frac{1}{{{\nu}}}\sum\limits_{k = 1}^K {\mathrm{tr}( {( {{{\bf{Y}}_k(\tau)} - {{\bf{G}}_k}{{\bf{X}}_k}} ){{( {{{\bf{Y}}_k}(\tau) - {{\bf{G}}_k}{{\bf{X}}_k}} )^H}} {{{( {{\bf{F}}_k^H{{\bf{F}}_k}} )^{ - 1}}}} } )} \label{GLRT Detector Derive 1}
\end{align}
where ${\bf Y}(\tau) = [{\bf Y}_1(\tau),{\bf Y}_2(\tau),\ldots,{\bf Y}_K(\tau)]$ and ${\bf G} = [{\bf G}_1,{\bf G}_2,\ldots,{\bf G}_K]$. It is easy to show that
\begin{align}
& \mathop {\max }\limits_{{\bf{G}},{\nu}} f( {{\bf{Y}}(\tau)|{\mathcal{H}_1},{\bf{G}},{\nu}} ) =  - L\sum\limits_{k = 1}^K {\ln \det ( {{\bf{F}}_k^H{{\bf{F}}_k}} )} - KL{N_{\rm{r}}} \nonumber \\
& - KL{N_{\rm{r}}}\ln \Bigg( {\frac{\pi }{{KL{N_{\rm{r}}}}}\sum\limits_{k = 1}^K {\mathrm{tr}( {( {{{\bf{Y}}_k(\tau)}{\bf{Y}}_k^H(\tau) - {{\bf{Y}}_k(\tau)}{\bf{X}}_k^H{{( {{{\bf{X}}_k}{\bf{X}}_k^H} )^{ - 1}}}{{\bf{X}}_k}{\bf{Y}}_k^H(\tau)} ){{( {{\bf{F}}_k^H{{\bf{F}}_k}} )^{ - 1}}}} )} } \Bigg) . \label{GLRT Detector Derive 5}
\end{align}
Similarly, under hypothesis $\mathcal{H}_0$, we can have
\begin{align}
\mathop {\max }\limits_{{\nu}} \ln f( {{\bf{Y}}(\tau)|{\mathcal{H}_0},{\nu}} ) &=  - L\sum\limits_{k = 1}^K {\ln \det ( {{\bf{F}}_k^H{{\bf{F}}_k}} )}  - KL{N_{\rm{r}}} \nonumber \\
&- KL{N_{\rm{r}}}\ln \Bigg( {\frac{\pi }{{KL{N_{\rm{r}}}}}\sum\limits_{k = 1}^K {\mathrm{tr}( {{{\bf{Y}}_k(\tau)}{\bf{Y}}_k^H(\tau){{( {{\bf{F}}_k^H{{\bf{F}}_k}} )^{ - 1}}}} )} } \Bigg) . \label{GLRT Detector Derive 8}
\end{align}
Finally, with (\ref{GLRT Detector Derive 5}) and (\ref{GLRT Detector Derive 8}), we can express (\ref{GLRT Detector Raw}) as
\begin{align}
T'(\tau) &= \mathop {\max }\limits_{{\bf{G}},{\nu}} \ln f( {{\bf{Y}}(\tau)|{\mathcal{H}_1},{\bf{G}},{\nu}} ) - \mathop {\max }\limits_{{\nu}} \ln f( {{\bf{Y}}(\tau)|{\mathcal{H}_0},{\nu}} ) \nonumber \\
&= - KL{N_{{\rm{r}}}} \ln \Bigg( {1 - \frac{{\sum\nolimits_{k = 1}^K {\mathrm{tr}( {{{\bf{Y}}_k(\tau)}{\bf{X}}_k^H{{( {{{\bf{X}}_k}{\bf{X}}_k^H} )^{ - 1}}}{{\bf{X}}_k}{\bf{Y}}_k^H(\tau){{( {{\bf{F}}_k^H{{\bf{F}}_k}} )^{ - 1}}}} )} }}{{\sum\nolimits_{k = 1}^K {\mathrm{tr}( {{{\bf{Y}}_k(\tau)}{\bf{Y}}_k^H(\tau){{( {{\bf{F}}_k^H{{\bf{F}}_k}} )^{ - 1}}}} )} }}} \Bigg) \overset{\mathcal{H}_1}{\underset{\mathcal{H}_0}{\gtrless}}  \gamma ' , \nonumber
\end{align}
and it is equivalent to (\ref{GLRT Detector}) with variable substitution $\gamma = 1 - \exp \big( { - \frac{{\gamma '}}{{KL{N_{\rm{r}}}}}} \big)$.

\section{Proof of Theorem 2 and 3}

First, we derive the probability of MD. Note that although we assume ${\bf X}_k$ is unitary in (\ref{Sync Signal Orthogonal}), the derivation below can also be applied to the more general non-unitary case. Define the following matrix
\begin{align}
{{\tilde{\bf{X}}_k}} &= {( {{{\bf{X}}_k}{\bf{X}}_k^H} )^{ - 1/2}}{{\bf{X}}_k} . \label{MD FA Probability Derive 1}
\end{align}
It can be verified that ${{\tilde{\bf{X}}}_k} \in \mathbb{C}^{{N_\mathrm{t}}\times L}$ satisfies
\begin{align}
{{\tilde{\bf{X}}}_k}\tilde{\bf{X}}_k^H = {( {{{\bf{X}}_k}{\bf{X}}_k^H} )^{ - 1/2}}{{\bf{X}}_k}{\bf{X}}_k^H{( {{{\bf{X}}_k}{\bf{X}}_k^H} )^{ - 1/2}} = {{\bf{I}}_{{N_{\rm{t}}}}} . \label{MD FA Probability Derive 2}
\end{align}
Then define another matrix ${\tilde{\bf{X}}_k^ \bot } \in \mathbb{C}^{{(L-N_\mathrm{t})}\times L}$ satisfying
\begin{align}
\Bigg[ {\begin{matrix}
{\tilde{\bf{ X}}_k^ \bot }\\
{{{\tilde{\bf{ X}}}_k}}
\end{matrix}} \Bigg]{\Bigg[ {\begin{matrix}
{\tilde{\bf{ X}}_k^ \bot }\\
{{{\tilde{\bf{ X}}}_k}}
\end{matrix}} \Bigg]^H} = \Bigg[ {\begin{matrix}
{{{\bf{I}}_{L - {N_{\rm{t}}}}}}&{\bf{0}}\\
{\bf{0}}&{{{\bf{I}}_{{N_{\rm{t}}}}}}
\end{matrix}} \Bigg] = {{\bf{I}}_L}, \label{MD FA Probability Derive 3}
\end{align}
i.e., ${\tilde{\bf{X}}_k^ \bot }$ and ${\tilde{\bf{X}}_k}$ constitute an $L \times L$ unitary matrix. With the property of unitary matrices, we can also write (\ref{MD FA Probability Derive 3}) as
\begin{align}
{\Bigg[ {\begin{matrix}
{\tilde{\bf{ X}}_k^ \bot }\\
{{{\tilde{\bf{ X}}}_k}}
\end{matrix}} \Bigg]^H}\Bigg[ {\begin{matrix}
{\tilde{\bf{ X}}_k^ \bot }\\
{{{\tilde{\bf{ X}}}_k}}
\end{matrix}} \Bigg] = {{\bf{I}}_L} . \label{MD FA Probability Derive 4}
\end{align}
With (\ref{Sync Signal Model}) and under hypothesis $\mathcal{H}_1$, we have
\begin{align}
{( {{\bf{F}}_k^H{{\bf{F}}_k}} )^{ - 1/2}}{{\bf{Y}}_k}(\tau) = {( {{\bf{F}}_k^H{{\bf{F}}_k}} )^{ - 1/2}}( {{{\bf{G}}_k}{{\bf{X}}_k} + {\bf{F}}_k^H{{\bf{Z}}_k}} ) = {{\tilde{\bf{ G}}}_k}{{\tilde{\bf{ X}}}_k} + {{\tilde{\bf{ Z}}}_k} \label{MD FA Probability Derive 5}
\end{align}
where the last equality is from (\ref{MD FA Probability Derive 1}), and
\begin{align}
{{\tilde{\bf{ G}}}_k} &= {( {{\bf{F}}_k^H{{\bf{F}}_k}} )^{ - 1/2}}{{\bf{G}}_k}{( {{{\bf{X}}_k}{\bf{X}}_k^H} )^{1/2}} \label{MD FA Probability Derive 6} \\
{{\tilde{\bf{ Z}}}_k} &= {( {{\bf{F}}_k^H{{\bf{F}}_k}} )^{ - 1/2}}{\bf{F}}_k^H{{\bf{Z}}_k}. \label{MD FA Probability Derive 7}
\end{align}
Since ${\bf Z}_k$ is with i.i.d. $\mathcal{CN}(0,\nu)$ entries, $\tilde{\bf Z}_k$ in (\ref{MD FA Probability Derive 7}) is also with i.i.d. $\mathcal{CN}(0,\nu)$ entries. Then we have
\begin{align}
&{( {{\bf{F}}_k^H{{\bf{F}}_k}} )^{ - 1/2}}{{\bf{Y}}_k(\tau)}{\bf{Y}}_k^H(\tau){( {{\bf{F}}_k^H{{\bf{F}}_k}} )^{ - 1/2}} = ( {{{\tilde{\bf{ G}}}_k}{{\tilde{\bf{ X}}}_k} + {{\tilde{\bf{ Z}}}_k}} ){\Bigg[ {\begin{matrix}
{\tilde{\bf{ X}}_k^ \bot }\\
{{{\tilde{\bf{ X}}}_k}}
\end{matrix}} \Bigg]^H}\Bigg[ {\begin{matrix}
{\tilde{\bf{ X}}_k^ \bot }\\
{{{\tilde{\bf{ X}}}_k}}
\end{matrix}} \Bigg]{( {{{\tilde{\bf{ G}}}_k}{{\tilde{\bf{ X}}}_k} + {{\tilde{\bf{ Z}}}_k}} )^H} \nonumber \\
&= [ {{{\tilde{\bf{ Z}}}_k}\tilde{\bf{ X}}_k^{ \bot H},{{\tilde{\bf{ G}}}_k} + {{\tilde{\bf{ Z}}}_k}\tilde{\bf{ X}}_k^H} ]{[ {{{\tilde{\bf{ Z}}}_k}\tilde{\bf{ X}}_k^{ \bot H},{{\tilde{\bf{ G}}}_k} + {{\tilde{\bf{ Z}}}_k}\tilde{\bf{ X}}_k^H} ]^H} = {{{\bf{ Z}}}_{k,1}}{\bf{ Z}}_{k,1}^H + ( {{{\tilde{\bf{ G}}}_k} + {{{\bf{ Z}}}_{k,2}}} ){( {{{\tilde{\bf{ G}}}_k} + {{{\bf{ Z}}}_{k,2}}} )^H} \label{MD FA Probability Derive 8}
\end{align}
where the first equality is from (\ref{MD FA Probability Derive 5}) and (\ref{MD FA Probability Derive 4}), the second equality is from (\ref{MD FA Probability Derive 3}), ${{{\bf{Z}}}_{k,1}} = {\tilde{\bf{Z}}_k}\tilde{\bf{X}}_k^{ \bot H} \in {\mathbb{C}^{{N_{\rm{r}}} \times \left( {L - {N_{\rm{t}}}} \right)}}$ and ${{{\bf{Z}}}_{k,2}} = {\tilde{\bf{Z}}_k}\tilde{\bf{X}}_k^{ H} \in {\mathbb{C}^{{N_{\rm{r}}} \times {N_{\rm{t}}}}}$ are both with i.i.d. $\mathcal{CN}(0,\nu)$ entries. In addition, we have
\begin{align}
&{( {{\bf{F}}_k^H{{\bf{F}}_k}} )^{ - 1/2}}{{\bf{Y}}_k(\tau)}{\bf{X}}_k^H{( {{{\bf{X}}_k}{\bf{X}}_k^H} )^{ - 1}}{{\bf{X}}_k}{\bf{Y}}_k^H(\tau){( {{\bf{F}}_k^H{{\bf{F}}_k}} )^{ - 1/2}} \nonumber \\
&= ( {{{\tilde{\bf{ G}}}_k}{{\tilde{\bf{ X}}}_k} + {{\tilde{\bf{ Z}}}_k}} )\tilde{\bf{ X}}_k^H{{\tilde{\bf{ X}}}_k}{( {{{\tilde{\bf{ G}}}_k}{{\tilde{\bf{ X}}}_k} + {{\tilde{\bf{ Z}}}_k}} )^H} = ( {{{\tilde{\bf{ G}}}_k} + {{{\bf{ Z}}}_{k,2}}} ){( {{{\tilde{\bf{ G}}}_k} + {{{\bf{ Z}}}_{k,2}}} )^H} \label{MD FA Probability Derive 9}
\end{align}
where the first equality is from (\ref{MD FA Probability Derive 5}) and (\ref{MD FA Probability Derive 1}), and the last equality is from (\ref{MD FA Probability Derive 2}). Therefore, with (\ref{MD FA Probability Derive 9}) and (\ref{MD FA Probability Derive 8}), the test statistic $T$ in (\ref{GLRT Detector}) under hypothesis $\mathcal{H}_1$ can be expressed as
\begin{align}
T(\tau)|\mathcal{H}_1 &= \frac{{\sum\nolimits_{k = 1}^K {\mathrm{tr}( {( {{{\tilde{\bf{ G}}}_k} + {{{\bf{ Z}}}_{k,2}}} ){{( {{{\tilde{\bf{ G}}}_k} + {{{\bf{ Z}}}_{k,2}}} )^H}}} )} }}{{\sum\nolimits_{k = 1}^K {\mathrm{tr}( {{{{\bf{ Z}}}_{k,1}}{\bf{ Z}}_{k,1}^H + ( {{{\tilde{\bf{ G}}}_k} + {{{\bf{ Z}}}_{k,2}}} ){{( {{{\tilde{\bf{ G}}}_k} + {{{\bf{ Z}}}_{k,2}}} )^H}}} )} }} \nonumber \\
&= \frac{{\sum\nolimits_{k = 1}^K {\| {{{(L/N_{\rm t})^{1/2}}}{( {{\bf{F}}_k^H{{\bf{F}}_k}} )^{ - 1/2}}{\bf{F}}_k^H{{\bf{H}}_k}{{\bf{W}}_k} + {{{\bf{ Z}}}_{k,2}}} \|_{\rm{F}}^2} }}{{\sum\nolimits_{k = 1}^K { ({\| {{{{\bf{ Z}}}_{k,1}}} \|_{\rm{F}}^2} + \| {{{(L/N_{\rm t})^{1/2}}}{( {{\bf{F}}_k^H{{\bf{F}}_k}} )^{ - 1/2}}{\bf{F}}_k^H{{\bf{H}}_k}{{\bf{W}}_k} + {{{\bf{ Z}}}_{k,2}}} \|_{\rm{F}}^2} )}}  \label{MD FA Probability Derive 10}
\end{align}
where the last equality is with (\ref{MD FA Probability Derive 6}) and (\ref{Sync Signal Orthogonal}). Note that in (\ref{MD FA Probability Derive 10}) it always holds that ${( {{\bf{F}}_k^H{{\bf{F}}_k}} )^{ - 1/2}}{\bf{F}}_k^H \cdot {\bf{F}}_k {( {{\bf{F}}_k^H{{\bf{F}}_k}} )^{ - 1/2}} = {\bf I}_{N_{\rm r}}$ for any ${\bf F}_k$ with full column rank. Hence we can rewrite (\ref{MD FA Probability Derive 10}) as
\begin{align}
T(\tau)|\mathcal{H}_1 =  \frac{{\sum\nolimits_{k = 1}^K {\| {{{(L/N_{\rm t})^{1/2}}}{\bf{F}}_k^H{{\bf{H}}_k}{{\bf{W}}_k} + {{{\bf{ Z}}}_{k,2}}} \|_{\rm{F}}^2} }}{{\sum\nolimits_{k = 1}^K { ({\| {{{{\bf{ Z}}}_{k,1}}} \|_{\rm{F}}^2} + \| {{{(L/N_{\rm t})^{1/2}}}{\bf{F}}_k^H{{\bf{H}}_k}{{\bf{W}}_k} + {{{\bf{ Z}}}_{k,2}}} \|_{\rm{F}}^2} )}} \label{MD FA Probability Derive 11}
\end{align}
for notational simplicity, where ${\bf F}_k$ satisfies ${\bf F}_k^H{\bf F}_k = {\bf I}_{N_\mathrm{r}}$. Since $T < \gamma$ is equivalent to $\frac{T}{1-T}<\frac{\gamma}{1-\gamma}$, the MD probability (\ref{MD Probability Definition}) can be expressed as (\ref{MD Probability}) according to (\ref{MD FA Probability Derive 11}).

To derive the FA probability, we can let ${\bf H}_k = {\bf 0}$ in (\ref{MD FA Probability Derive 11}), yielding the test statistic $T$ under hypothesis $\mathcal{H}_0$
\begin{align}
T|\mathcal{H}_0 = \frac{{\sum\nolimits_{k = 1}^K {\| {{{{\bf{ Z}}}_{k,2}}} \|_{\rm{F}}^2} }}{{ \sum\nolimits_{k = 1}^K ( {\| {{{{\bf{ Z}}}_{k,1}}} \|_{\rm{F}}^2} + {\| {{{{\bf{ Z}}}_{k,2}}} \|_{\rm{F}}^2} ) }} . \nonumber
\end{align}
Therefore, the FA probability (\ref{FA Probability Definition}) can be expressed as (\ref{FA Probability}).

\section{Proof of Lemma 1}

First, we derive the characteristic function (CF) and the $n$th non-central moment of $X$, which will be used latter. For $x_m \sim \mathcal{CN}(0,\lambda_m)$, ${X_m} = {| {{x_m}} |^2}$ follows exponential distribution and the CF of $X_m$ is
\begin{align}
{\varphi _{{X_m}}}( \omega  ) = \mathbb{E}\{ {{e^{ - j\omega {X_m}}}} \} = \frac{1}{{1 + j{\lambda _m}\omega }} . \nonumber
\end{align}
Then, the CF of $X = \sum\nolimits_{m = 1}^M {{X_m}} $ is
\begin{align}
{\varphi _X}( \omega  ) = \mathbb{E}\{ {{e^{ - j\omega {X}}}} \} = \prod\limits_{m = 1}^M {{\varphi _{{X_m}}}( \omega  )} = \prod\limits_{m = 1}^M {\frac{1}{{1 + j{\lambda _m}\omega }}} . \label{Chi-Square Distribution CF}
\end{align}
Let $f_X(x)$ denote the PDF of $X$. According to the relation between CF and PDF
\begin{align}
{\varphi _X}( \omega  ) = \int_{0}^\infty  {{f_X}( x ){e^{ - j\omega x}}\mathrm{d}x}, \label{Chi-Square Distribution Derive 1}
\end{align}
we have
\begin{align}
\frac{{{\mathrm{d}^n}{\varphi _X}( \omega  )}}{{\mathrm{d}{\omega ^n}}} = {( { - j} )^n}\int_{0}^\infty  {{x^n}{f_X}( x ){e^{ - j\omega x}}\mathrm{d}x} . \nonumber
\end{align}
Therefore,
\begin{align}
\mathbb{E}\{ {{X^n}} \} = \int_{0}^\infty  {{x^n}{f_X}( x )\mathrm{d}x} = {(-j)^{-n}} \frac{{{\mathrm{d}^n}{\varphi _X}( \omega  )}}{{\mathrm{d}{\omega ^n}}}{\bigg|_{\omega  = 0}} . \label{Chi-Square Distribution Derive 2}
\end{align}
With the general Leibniz rule, the $n$th derivative of ${\varphi _X}( \omega  )$ is
\begin{align}
\frac{{{\mathrm{d}^n}{\varphi _X}( \omega  )}}{{\mathrm{d}{\omega ^n}}} &= \frac{{{\mathrm{d}^n}}}{{\mathrm{d}{\omega ^n}}}\Bigg( {\prod\limits_{m = 1}^M {\frac{1}{{1 + j{\lambda _m}\omega }}} } \Bigg) \nonumber \\
&= \sum\limits_{{k_1} + {k_2} + \cdots  + {k_M} = n} { \binom{n}{k_1,k_2,\ldots,k_M} \prod\limits_{m = 1}^M { \frac{{{\mathrm{d}^{{k_m}}}}}{{\mathrm{d}{\omega ^{{k_m}}}}}\bigg( {\frac{1}{{1 + j{\lambda _m}\omega }}} \bigg)  }} \nonumber \\
&= \sum\limits_{{k_1} + {k_2} + \cdots  + {k_M} = n} { \frac{{n!}}{{{k_1}!{k_2}! \cdots {k_M}!}} \prod\limits_{m = 1}^M {\frac{{{{( { - j{\lambda _m}} )^{{k_m}}}}{{k_m}}!}}{{{{( {1 + j{\lambda _m}\omega } )^{{k_m} + 1}}}}}} } \nonumber \\
&= {( { - j} )^n} n!\sum\limits_{{k_1} + {k_2} + \cdots  + {k_M} = n} {\prod\limits_{m = 1}^M {\frac{{\lambda _m^{{k_m}}}}{{{{( {1 + j{\lambda _m}\omega } )^{{k_m} + 1}}}}}} } \label{Chi-Square Distribution Derive 3}
\end{align}
where each $k_m$ is a non-negative integer. Substituting (\ref{Chi-Square Distribution Derive 3}) into (\ref{Chi-Square Distribution Derive 2}) yields
\begin{align}
\mathbb{E} \{ {{X^n}} \} = n!\sum\limits_{{k_1} + {k_2} + \cdots  + {k_M} = n} { \prod \limits_{m = 1}^M {\lambda _m^{{k_m}}} } . \label{Chi-Square Distribution Moment}
\end{align}

Then, we derive the asymptotic value for the CDF of $X/Y$. Since $X$ and $Y$ are independent with each other, we have
\begin{align}
\mathbb{P}\bigg\{ {\frac{X}{Y} < t} \bigg\} = \mathbb{P}\{ {X < tY} \} = \int_{0}^\infty  {{f_Y}( y )} \int_{0}^{ty} {{f_X}( x )\mathrm{d}x} \mathrm{d}y. \label{Generalized F Distribution CDF Derive 1}
\end{align}
To derive the asymptotic value of (\ref{Generalized F Distribution CDF Derive 1}) when $t$ is small, we use (\ref{Chi-Square Distribution CF}) to obtain the Taylor series expansion of ${\varphi _X}( \omega  )$ at $1/(j\omega)=0$
\begin{align}
{\varphi _X}( \omega  )  = {{\frac{1}{{(j\omega)^M }}} }\prod\limits_{m = 1}^M {\frac{1}{{1/j\omega  + {\lambda _m}}}}  = {{\frac{1}{{(j\omega)^M }}} }\sum\limits_{k = 0}^\infty  {{{ {\frac{a_k}{{(j\omega)^k }}} }}} \nonumber
\end{align}
where
\begin{align}
{a_k} = \sum\limits_{{k_1} + {k_2} \cdots  + {k_M} = k} {\frac{1}{{{k_1}!{k_2}! \cdots {k_M}!}}\prod\limits_{m = 1}^M {\frac{{{{( { - 1} )^{{k_m}}}}}}{{\lambda _m^{{k_m} + 1}}}} }  . \label{Generalized F Distribution CDF Derive 6}
\end{align}
Then the PDF of $X$ is
\begin{align}
{f_X}( x ) &= \frac{1}{{2\pi }}\int_{ - \infty }^\infty  {{\varphi _X}( \omega  ){e^{j\omega x}}\mathrm{d}\omega } = \frac{1}{{2\pi }}\int_{ - \infty }^\infty  {{{ {\frac{1}{{(j\omega)^M }}} }}\sum\limits_{k = 0}^\infty  {{{ {\frac{a_k}{{(j\omega)^k }}} }}} {e^{j\omega x}}\mathrm{d}\omega } = u( x ) \cdot \sum\limits_{k = 0}^\infty  {\frac{{{a_k}{x^{M + k - 1}}}}{{( {M + k - 1} )!}}} \nonumber
\end{align}
where $u(x)$ denotes the unit step function, i.e., $u(x)=1$ when $x\geq0$ and $u(x)=0$ when $x<0$.
Then we have
\begin{align}
\int_{-\infty}^{ty} {{f_X}( x )\mathrm{d}x}  = \int_0^{ty} {\sum\limits_{k = 0}^\infty  {\frac{{{a_k}{x^{M + k - 1}}}}{{( {M + k - 1} )!}}} \mathrm{d}x}  = \sum\limits_{k = 0}^\infty  {\frac{{{a_k}{{( {ty} )^{M + k}}}}}{{( {M + k} )!}}} \nonumber
\end{align}
and hence
\begin{align}
\mathbb{P}\bigg\{ {\frac{X}{Y} < t} \bigg\} &= \int_{0}^\infty  {{f_Y}( y )} \int_{0}^{ty} {\sum\limits_{k = 0}^\infty  {\frac{{{a_k}{x^{M + k - 1}}}}{{( {M + k - 1} )!}}}\mathrm{d}x} \mathrm{d}y = \int_{-\infty}^\infty  {{f_Y}( y )\sum\limits_{k = 0}^\infty  {\frac{{{a_k}{{( {ty} )^{M + k}}}}}{{( {M + k} )!}}} \mathrm{d}y} \nonumber \\
&= \sum\limits_{k = 0}^\infty  {\frac{{{a_k}{t^{M + k}}\mathbb{E}\{ {{y^{M + k}}} \}}}{{( {M + k} )!}}} = \sum\limits_{k = 0}^\infty  {{a_k}{t^{M + k}}\sum\limits_{{l_1} + {l_2} +  \cdots  + {l_N} = M + k} { \prod \limits_{n=1}^{N} \sigma _n^{{l_n}}} } \nonumber \\
&\approx {{a_0}{t^{M}}\sum\limits_{{l_1} + {l_2} +  \cdots  + {l_N} = M} { \prod \limits_{n=1}^{N} \sigma _n^{{l_n}}} }, \;\; \text{when } t \text{ is small}, \label{Generalized F Distribution CDF Derive 7}
\end{align}
where the last equality is with (\ref{Chi-Square Distribution Moment}). Substituting (\ref{Generalized F Distribution CDF Derive 6}) with $k=0$ into (\ref{Generalized F Distribution CDF Derive 7}) yields (\ref{Generalized F Distribution CDF Asymptotic}).


\end{document}